\documentclass[fleqn,10pt]{wlscirep}
\usepackage{amssymb}
\usepackage{algorithm,algorithmic}
\usepackage{booktabs}
\usepackage{threeparttable}
\usepackage{multirow}
\usepackage{url}
\usepackage{amssymb}
\usepackage{color}
\usepackage{caption,subcaption}
\usepackage{graphicx}
\usepackage{latexsym}

\title{Evolving nature of human contact networks with its impact on epidemic processes}

\author[1,2]{Cong Li}
\author[1]{Jing Li}
\author[1,2,*]{Xiang Li}
\affil[1]{Adaptive Networks and Control Lab, Department of Electronic Engineering, Fudan University, 200433, China}
\affil[2]{Research Center of Smart Networks and Systems, School of Information Science and Engineering, Fudan University, 200433, China}

\affil[*]{lix@fudan.edu.cn}


\begin{abstract}
Human contact networks are constituted by a multitude of individuals and pairwise contacts among them. However, the dynamic nature, which generates the evolution of human contact networks, of contact patterns is not known yet. Here, we analyse three empirical datasets and identify two crucial mechanisms of the evolution of temporal human contact networks, \emph{i.e.} the activity state transition laws for an individual to be socially active, and the contact establishment mechanism that active individuals adopt. We consider both of the two mechanisms to propose a temporal network model, named the memory driven (MD) model, of human contact networks. Then we study the susceptible-infected (SI) spreading processes on empirical human contact networks and four corresponding temporal network models, and compare the full prevalence time of SI processes with various infection rates on the networks. The full prevalence time of SI processes in the MD model is the same as that in real-world human contact networks. Moreover, we find that the individual activity state transition promotes the spreading process, while, the contact establishment of active individuals suppress the prevalence. Apart from this, we observe that even a small percentage of individuals to explore new social ties is able to induce an explosive spreading on networks. The proposed temporal network framework could help the further study of dynamic processes in temporal human contact networks, and offer new insights to predict and control the diffusion processes on networks.
\end{abstract}

\begin{document}

\flushbottom
\maketitle
%
%
\thispagestyle{empty}


\section*{Introduction}
The evolution nature of human interaction creates diverse temporal properties, which fundamentally influence the epidemic spreading. Interactions may be characterized and modeled by the human interaction networks \cite{granovetter1973strength,krackhardt2003strength,starnini2013modeling,Zhang2015Human,zhang2015characterizing}, where a node represents a person and the interaction between two persons is a link. In 2004, Pentland \emph{et al.} from MIT Media Laboratory took the lead in tracking and recording human interactions with personal mobile phones, and proposed a stochastic process model to capture the co-evolution of social relationships and individual behaviours \cite{Nathan2006Reality,dong2011modeling}. Many previous studies of interaction networks assume that the interactions are fixed or the spreading processes are much faster than the evolution of networks. Nowadays, with the rapid advances in technology, high quality and time-resolved datasets of human behaviours can be easily obtained \cite{Cattuto2010Dynamics,isella2011s,saramaki2015seconds,saramaki2014persistence,sekara2016fundamental} and have provided an unprecedented opportunity to better understand of human society dynamics with temporal human interaction networks \cite{ubaldi2016asymptotic,ubaldi2017burstiness,Medus2014Memory}. However, it remains an important challenge to  unveil the mechanisms driving the evolution of human interaction networks, and to further capture the effects of such mechanisms on dynamical processes occurring on the network \cite{barabasi2005origin,karsai2011small,Rocha2013Bursts,miritello2013limited,vestergaard2014memory,Scholtes2014Causality}.

For human interaction networks, in general, each agent at time $t$ is either active, contacting with other agents, or inactive, isolated. The activity state, \emph{i.e.} active or inactive of an agent and the contacts between pairwise agents alter with time. Rocha and Blondel \cite{Rocha2013Bursts} introduced a temporal network model, where the time interval between two activity states of each agent follows a stochastic process with a certain distribution.
Differently, some researchers consider that the nodes in the inactive state are not totally isolated, they can still receive contacts from other active nodes. A typical example is the framework of activity driven networks \cite{perra2012activity,hoppe2013mutual,pozzana2017epidemic,alessandretti2017random}.
Perra \emph{et al.} \cite{perra2012activity} proposed the original activity driven model of temporal networks, in which an agent is either active with a probability $p$ or inactive with a probability $1-p$ at each unit time. Since the active probability of an agent is decided by its activity potential extracted from a given distribution, regardless of its status at last time step, the activity driven model generates a sequence of memoryless networks. Besides, the social contacts between active nodes and other nodes are randomly established. A further extension of this model reduces the randomness of link establishments by introducing a mutual selection mechanism, where the destination of a contact depends mutually on the activity potential of the agents on both ends of that connection \cite{hoppe2013mutual}. Similarly, some studies add a time-invariant quantity, namely attractiveness for each agent, accounting for the fact that some agents are more likely to be connected by other agents when building a contact \cite{pozzana2017epidemic,alessandretti2017random}.

Most previous studies assume that the transition of activity states and the building of social interactions are both stochastic processes \cite{Rocha2013Bursts,perra2012activity,hoppe2013mutual,pozzana2017epidemic,alessandretti2017random}. However, a wealth of empirical observations have shown that the mechanisms which govern the evolution of human contact networks are far from random \cite{ubaldi2016asymptotic,ubaldi2017burstiness,Medus2014Memory,miritello2013limited,karsai2014time,kim2015scaling,valdano2015predicting,jing2018quantifying}.

Individuals often remember the agents whom they previously interacted with, and thus form their own social circles \cite{miritello2013limited}. Besides, due to the continuity of social interactions, the activity state of an agent is unlikely to be altered in a sudden \cite{sekara2016fundamental}, especially when the agent is actively engaging in a social event with a lot of peers. Valdano \emph{et al.} \cite{valdano2015predicting} proposed a homogeneous assumption that the transition probability between the active state and inactive state for each agent is identically assigned with the same value. Although the assumption partially captures the evolution of the activity state, an explanation of transition mechanism is still missing. When it comes to the evolution of contacts between agents, the effect of memory, \emph{i.e.} an agent preserving their pervious contacts with a certain probability, should be considered within the temporal network model. The model introduced in \cite{vestergaard2014memory} incorporates four distinct memory mechanisms to describe the rates of creation and disappearance of contacts in empirical networks. The authors consider the rates follow a power-law decaying functional forms, akin to a rich-get-richer preferential attachment mechanism \cite{barabasi1999emergence}. Recently, more realistic mechanisms based on the activity-driven framework have been proposed \cite{ubaldi2016asymptotic,ubaldi2017burstiness,karsai2014time,kim2015scaling}. They consider a non-Markovian reinforcement process, in which agents are more inclined to allocate their contacts towards already existing social ties rather than create new relationships. In particular, the probability of an agent exploring a new social relationship is determined by its cumulative degree, \emph{i.e.} the number of distinct agents contacted during observation period, however, the transition of the activity states of each agent is not taken into account.

The aforementioned studies have investigated various topological and temporal features of real social systems from different perspectives. However, a better understanding of the mechanisms driving the evolution of the temporal social contact networks, and the effects of network evolution on dynamic processes, such as epidemic spreading \cite{Romualdo2001Epidemic,Castellano2010Thresholds,Wang2016Identifying,Qu2017Ranking}, information propagation \cite{Scholtes2014Causality,Iribarren2009Impact} and innovation diffusion \cite{Montanari2010The,kreindler2014rapid}, is still urgently needed.

Exploring several datasets of human contact networks, we report two generic mechanisms of the evolution of real networks: (1) the transition probability of activity states is a function of the degree of agents, (2) the time interval between the reconnection of node pairs follows a power-law distribution. Based on the two mechanisms, we propose a temporal network model, namely the memory driven model, to characterize the evolution of social networks. The memory here refers that an agent inclines to concentrated towards recently contacted partners. With the new framework, we study the epidemic spreading processes in both synthetic and real temporal networks. The full prevalence time of Susceptible-Infected (SI) spreading in the memory driven model is the same as that in real networks. Moreover, we find that the two mechanisms of the network evolution have distinct effects on the epidemic spreading on real networks, that is, the activity state transition promotes the spreading process, however, the contact memory hampers the spreading. Apart from this, we demonstrate that even a smaller fraction of interactions connecting to new social relationships is able to induce an explosive spreading on networks.

This paper is organized as follows. In Section II, we describe three datasets of real-world human interactive activities, and introduce the definitions of human contact networks. In Section III, we analyse the empirical datasets to obtain statistic properties of human contact networks and propose a temporal network model.
Section IV focuses on the epidemic spreading processes on the synthetic networks and real networks. Finally, we conclude this work in Section V.

\section*{Empirical Datasets and definitions}
\subsection*{Dataset description}
The three datasets of offline human interactive activities include the offline sex interaction dataset ``Sex6yr", the physical proximity interaction datasets ``MIT\_RM" and ``School". The ``Sex6yr" dataset collects the real offline sex trading activities among sellers and buyers. The ``MIT\_RM" dataset records the physical proximity interactions among MIT students via the bluetooth device in mobile phones. The ``School" dataset collects the face-to-face proximity interactions among teachers and students in a French school by the Radio-Frequency IDentification (RFID) device embedded in badges. The contact records of each dataset are in the same format $(i,j,t)$, which represents a contact between two individuals $i$ and $j$ started at time $t$. Details of the datasets are introduced in Appendix A and summarized in Table \ref{dataset}.

\renewcommand{\arraystretch}{1.5}
\begin{table*}[]
\centering
\setlength{\abovecaptionskip}{0.cm}
\setlength{\belowcaptionskip}{-0.cm}
\caption{Description of the three datasets, where $T$ is the time span of contact sequences, $\triangle t$ is the resolution (time interval between two records in time order), $N$ is the number of agents, $L_C$ is the number of distinct contacts, and $R$ is the number of records.} \label{dataset}
\begin{tabular*}{1.0\textwidth}{@{\extracolsep{\fill}}c c c c c c}
\hline
Dataset & $T$ & $\triangle t$ & $N$ & $L_C$  & $R$\\
\hline
Sex6yr  & 2,232 day &  1 day &  16,102 & 38,995 & 50,185  \\
MIT\_RM & 232 day & 6 min & 96 & 2,539 & 55,306 \\
School  & 8.67 hour & 20 sec & 236 & 5,901 & 37,402 \\
\hline
\end{tabular*}
\end{table*}

\begin{figure}[t]
\centering
\includegraphics[width=0.6\textwidth]{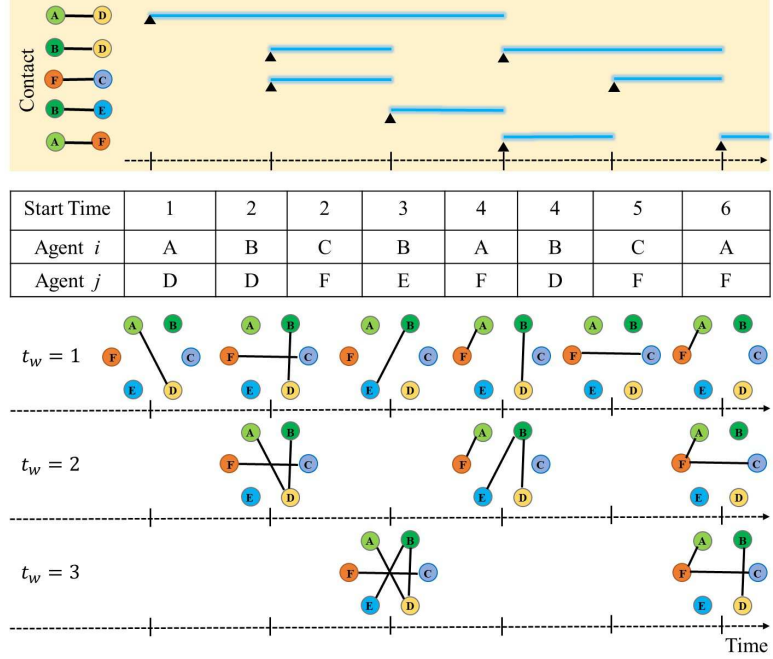}
\caption{Schematic illustration of the construction of temporal networks and the time slice of a contact sequence, where $t_w$ is the length of a time step.}
\label{network}
\end{figure}
\subsection*{Definitions}
\textbf{Definition 1} (Human contact networks).
Human contact network is a kind of temporal network, which consists of the evolution of network topology with time. We segment the empirical data into adjacency time steps of length $t_{w}=w\triangle t$, where $\triangle t$ is the resolution, $w$ is the number of resolutions in a time step. A human contact network is a sequence of networks in time order $\mathcal{G}(t_w) = \{G_1, G_2, ..., G_t, ..., G_T\}$, where $G_t = (\mathcal{N}_t, \mathcal{L}_t)$ is the network taking place at time step $t$ over time window $[(t-1)t_w, t\cdot t_w)$, and $T$ is the total number of time steps. An illustration of the human contact network is shown in Figure \ref{network}. In this work, we select $t_{w}$ as 1 month, 1 day and 5 mins for ``Sex6yr", ``MIT\_RM" and ``School", respectively (explanation for the selection is given in Appendix B). The aggregated network of $\mathcal{G}(t_w)$ is $\mathcal{G}_{T}=\bigcup_{t=1}^{t=T}G_{t}=(\mathcal{N}, \mathcal{L})$, where $\mathcal{N}=\bigcup_{t=1}^{t=T}\mathcal{N}_{t}$ and $\mathcal{L}=\bigcup_{t=1}^{t=T}\mathcal{L}_{t}$. We denote the number of agents and the number of contacts by $N=|\mathcal{N}|$ and $L_C=|\mathcal{L}|$, respectively.

\section*{Analysis of Temporal Properties}
Before we introduce the temporal network model, we first empirically analyse statistic properties of human contact networks, \emph{i.e.} the activity state transition of individuals and the contact memory of node pairs, which are the fundamental mechanisms for the design of our temporal network model.

\subsection*{Individual activity state transition}
We here assume that each node $i$ at time step $t$ in the network is in one of two states: active state, connecting with other nodes, denoted by $a$ and inactive state ${\i}$. The sets of nodes in active state and in inactive state at time step $t$ are denoted by $\mathcal{N}_{a}(t)$ and $\mathcal{N}_{\i}(t)$, respectively. We here study whether the activity state of an individual $i$ at time step $t+1$ depends on the state at previous step $t$, and how the degree $k$ of individual $i$ at time step $t$ influence the activity state transition. We denote the number of nodes with degree $k$ in active state at time step $t$ by $N_{k,a}(t)=|\mathcal{N}_{k,a}(t)|$, and the number of nodes with degree $k$ in inactive state at time step $t$ by $N_{k,{\i}}(t)=|\mathcal{N}_{k,{\i}}(t)|$, where $|\mathcal{N}|$ represents the number of nodes in a node set $\mathcal{N}$. The transition probability of nodes with degree $k$ transferring from active state $a$ to inactive state ${\i}$ at one time step is
\begin{equation}
P_{AI}(k)=\frac{\sum_{t=1}^{T-1} |\mathcal{N}_{k,a}(t)\bigcap \mathcal{N}_{\i}(t+1)|}{\sum_{t=1}^{T-1} N_{k,a}(t)},
\end{equation}

The transition probability of nodes with degree $k$ staying at active state $a$ in one time step is $P_{AA}(k) = 1-P_{AI}(k)$. Note that the degree of an individual might change over time, since the individual might take part in or leaves from the interactions of the system across time.
\begin{figure}[!ht]
\setlength{\abovecaptionskip}{0.cm}
\setlength{\belowcaptionskip}{-0.cm}
\centering
\begin{subfigure}[t]{0.35\textwidth}
\centering
\setlength{\abovecaptionskip}{0.cm}
\setlength{\belowcaptionskip}{-0.cm}
\includegraphics[width=1\textwidth]{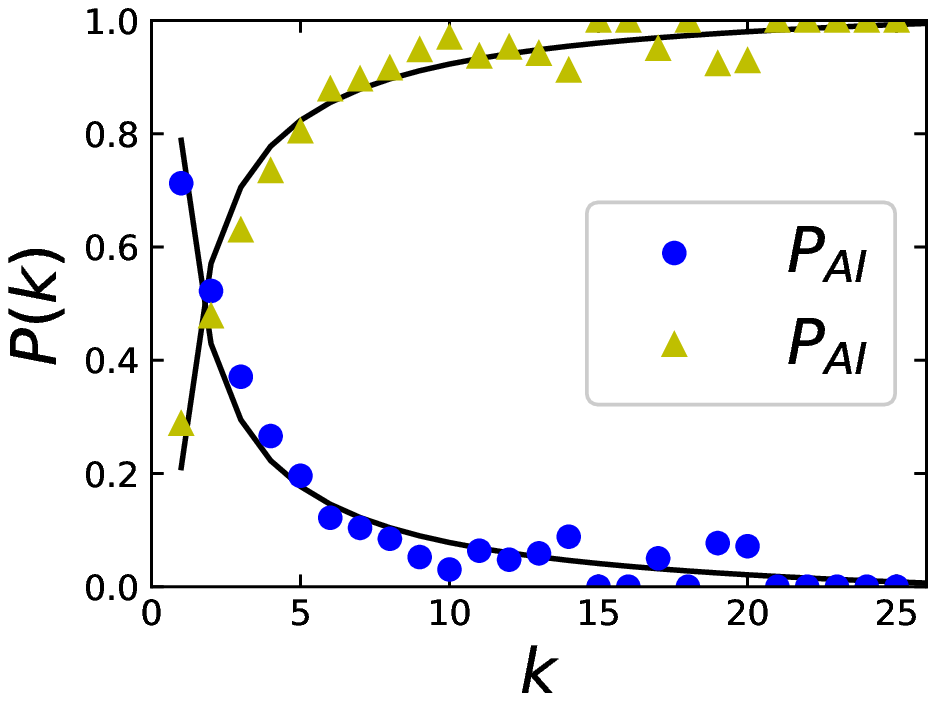}
\subcaption{}
\end{subfigure}
\begin{subfigure}[t]{0.35\textwidth}
\centering
\setlength{\abovecaptionskip}{0.cm}
\setlength{\belowcaptionskip}{-0.cm}
\includegraphics[width=1\textwidth]{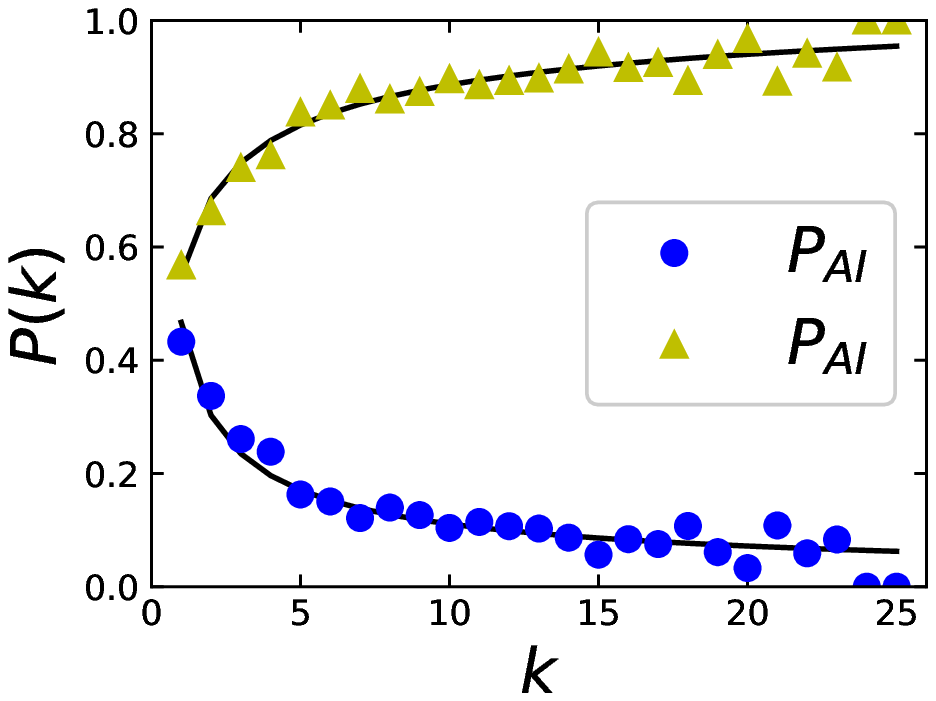}
\subcaption{}
\end{subfigure}
\begin{subfigure}[t]{0.35\textwidth}
\centering
\setlength{\abovecaptionskip}{0.cm}
\setlength{\belowcaptionskip}{-0.cm}
\includegraphics[width=1\textwidth]{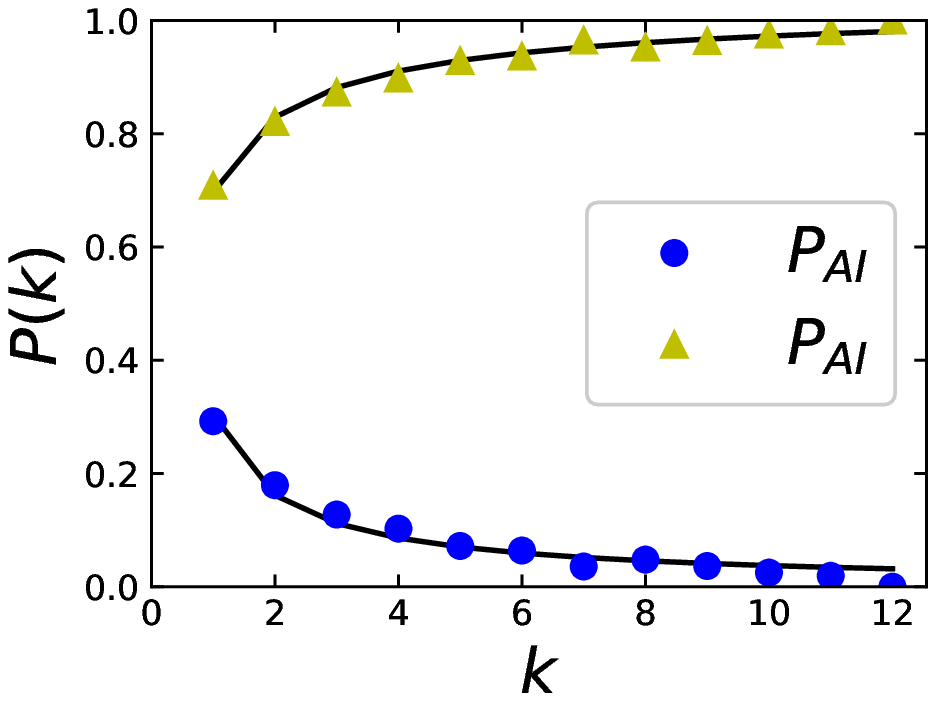}
\subcaption{}
\end{subfigure}
\begin{subfigure}[t]{0.35\textwidth}
\centering
\setlength{\abovecaptionskip}{0.cm}
\setlength{\belowcaptionskip}{-0.cm}
\includegraphics[width=1\textwidth]{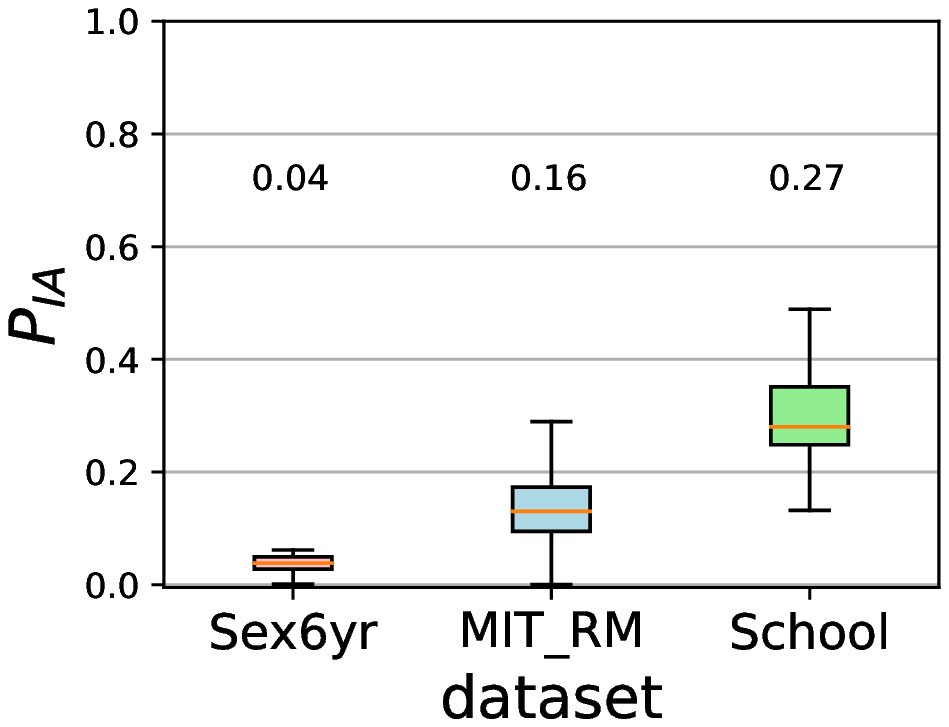}
\subcaption{}
\end{subfigure}
\caption{Transition probabilities between active state and inactive state as functions of the current degree $k$ of the node, for the three empirical datasets. Solid lines represent the fitting curves: (a) $P_{AI}(k)=0.78k^{-0.84}$ (Sex6yr), (b) $P_{AI}(k)=0.45k^{-0.56}$ (MIT\_RM) and (c) $P_{AI}(k)=0.33k^{-0.91}$ (School). (d) The firing rate $P_{IA}$ in the empirical datasets. The average firing rate $P_{IA}$ is 0.04 (Sex6yr), 0.16 (MIT\_RM), or 0.27 (School).}
\label{transfer}
\end{figure}

The transition probabilities characterize the relation between the activity state of node $i$ with degree $k$ at the current time step and that at next time step. We study how the transition probability, \emph{i.e.} $P_{AI}(k)$ or $P_{AA}(k)$, varies with the degree $k$. Figs. \ref{transfer}(a)-\ref{transfer}(c) show the transition probabilities of remaining the active status or switching to the inactive status calculated on three empirical networks. Remarkably, the transition probability $P_{AI}(k)$ or $P_{AA}(k)$ is as a power-law function of degree $k$. The greater the degree of active agents at the current time step, the smaller the probability they become inactive at the next time step. That is to say, agents with high social activity (high degree) have more tendency to maintain the active status all the time. However, we cannot find the relation between the transition probability $P_{IA}(k)$ (or $P_{II}(k)$) and degree $k$, since the degree of nodes in inactive state is 0. Hence, we study the transition property of inactive nodes with the firing rate $b=\frac{|\mathcal{N}_{\i}(t)\bigcap \mathcal{N}_{a}(t+1)|}{N_{\i}(t)}$, which is the probability of an inactive node becoming active at one time step. The average firing rate $P_{IA}$ is expressed as
\begin{equation}
P_{IA}=\frac{1}{T-1}\sum_{t=1}^{T-1} \frac{|\mathcal{N}_{\i}(t)\bigcap \mathcal{N}_{a}(t+1)|}{N_{\i}(t)}.
\end{equation}
Fig. \ref{transfer}(d) demonstrates that the variance of firing rates at different time steps is small.

\subsection*{Contact establishment of active individuals}
\begin{figure}[!ht]
\centering
\begin{subfigure}[t]{0.239\textwidth}
\centering
\setlength{\abovecaptionskip}{0.cm}
\setlength{\belowcaptionskip}{-0.cm}
\includegraphics[width=1\textwidth]{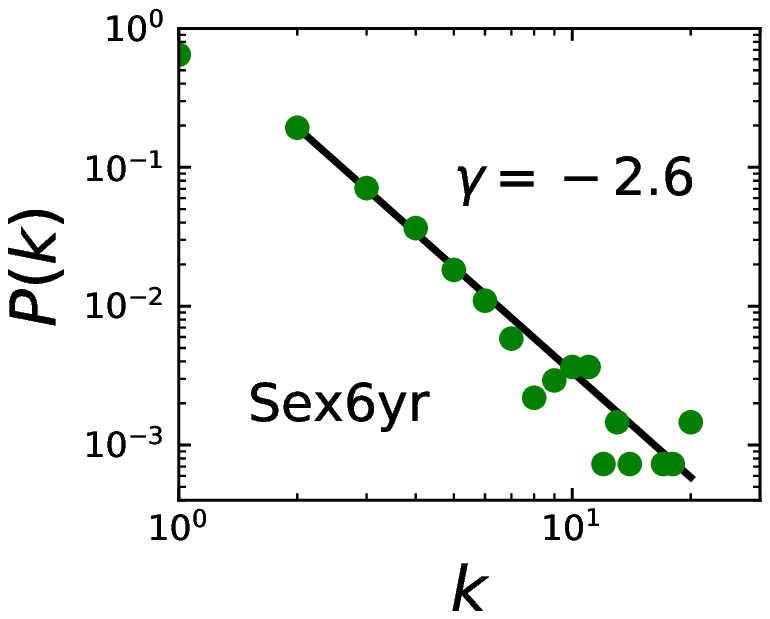}
\subcaption{}
\end{subfigure}
\begin{subfigure}[t]{0.249\textwidth}
\centering
\setlength{\abovecaptionskip}{0.cm}
\setlength{\belowcaptionskip}{-0.cm}
\includegraphics[width=1\textwidth]{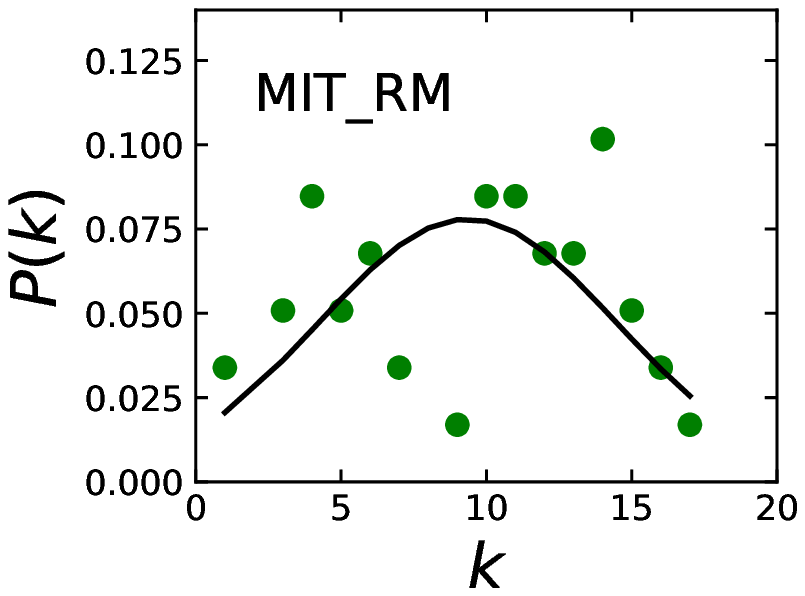}
\subcaption{}
\end{subfigure}
\begin{subfigure}[t]{0.241\textwidth}
\centering
\setlength{\abovecaptionskip}{0.cm}
\setlength{\belowcaptionskip}{-0.cm}
\includegraphics[width=1\textwidth]{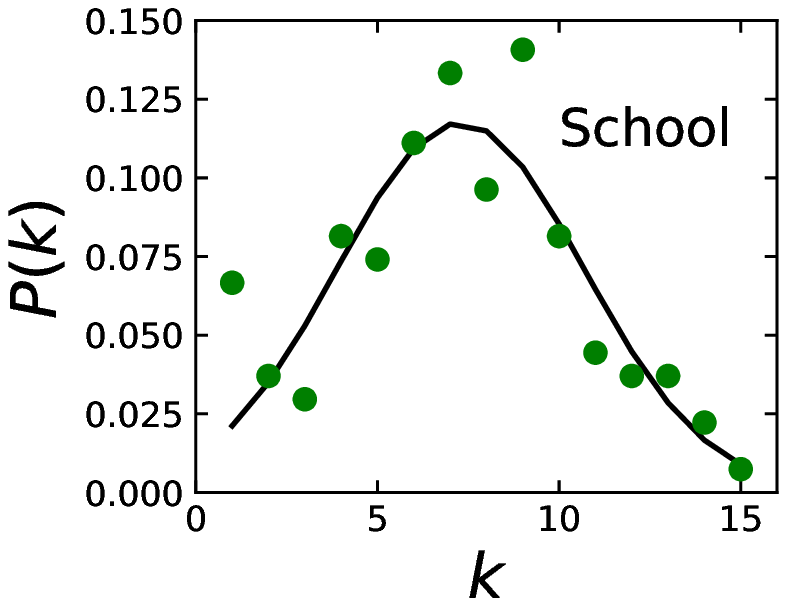}
\subcaption{}
\end{subfigure}
\begin{subfigure}[t]{0.23\textwidth}
\centering
\setlength{\abovecaptionskip}{0.cm}
\setlength{\belowcaptionskip}{-0.cm}
\includegraphics[width=1\textwidth]{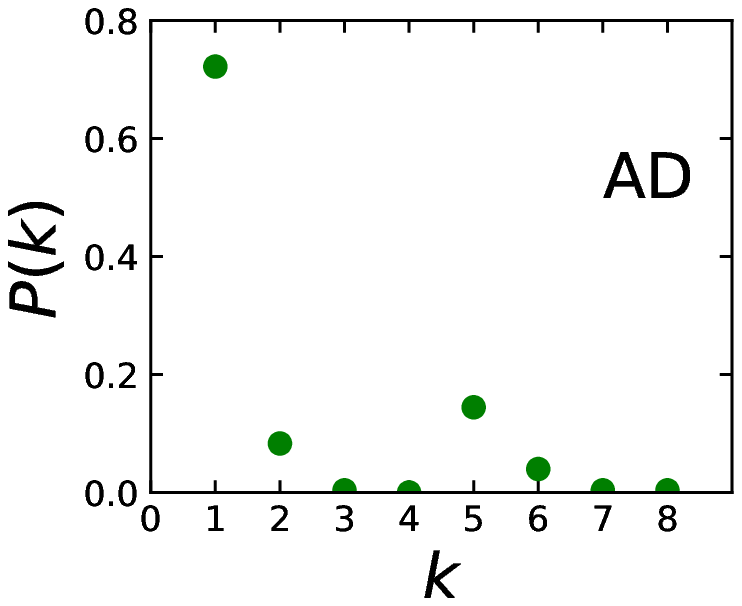}
\subcaption{}
\end{subfigure}
\caption{(a)(b)(c) Degree distributions $P(k)$ at each time step in three empirical datasets and (d) Degree distribution $P(k)$ at each time step in activity driven (AD) model \cite{perra2012activity} with $s=5$, $N=1000$, and node activity $a$ is sampled from $F(a)\propto a^{-\nu}$ with $\nu=2.1$, $10^{-2}\leq a\leq1$. }
\label{pdf}
\end{figure}

Most previous studies \cite{perra2012activity,hoppe2013mutual,pozzana2017epidemic,alessandretti2017random} assume that all active individuals contact with $s$ other individuals at each time step, where $s=m$ is a constant. However, we find that the number of contacts (degree) at each time step follows a different distribution for different empirical datasets. Fig. \ref{pdf} suggests that the assumption of the same number of contacts cannot reflect the real properties, thus, the distribution $F(s)$ of contact number $s$ should be considered for the temporal network models.

\begin{figure}[!ht]
\centering
\begin{subfigure}[t]{0.3\textwidth}
\centering
\setlength{\abovecaptionskip}{0.cm}
\setlength{\belowcaptionskip}{-0.cm}
\includegraphics[width=1\textwidth]{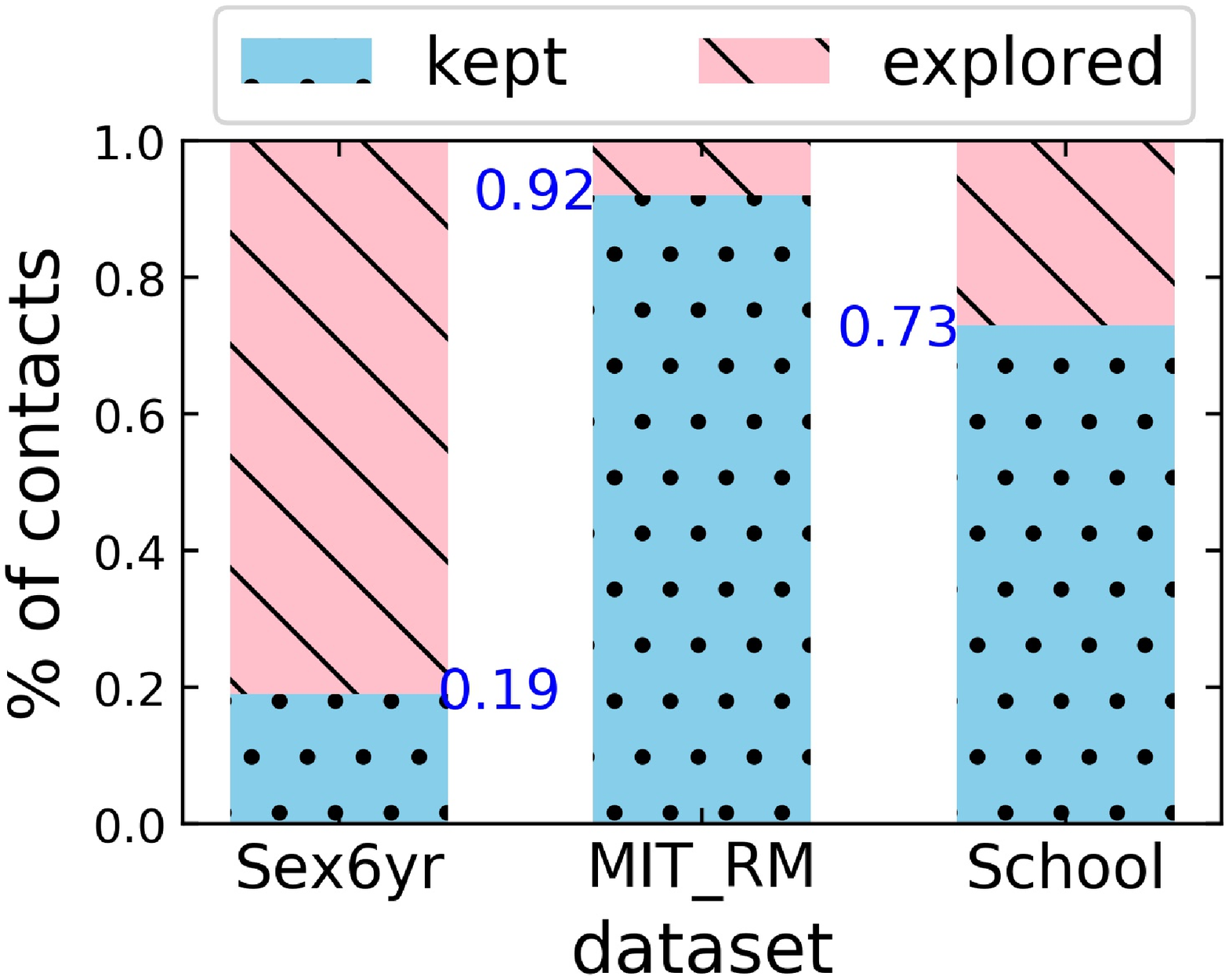}
\subcaption{}
\end{subfigure}
\begin{subfigure}[t]{0.3\textwidth}
\centering
\setlength{\abovecaptionskip}{0.cm}
\setlength{\belowcaptionskip}{-0.cm}
\includegraphics[width=1\textwidth]{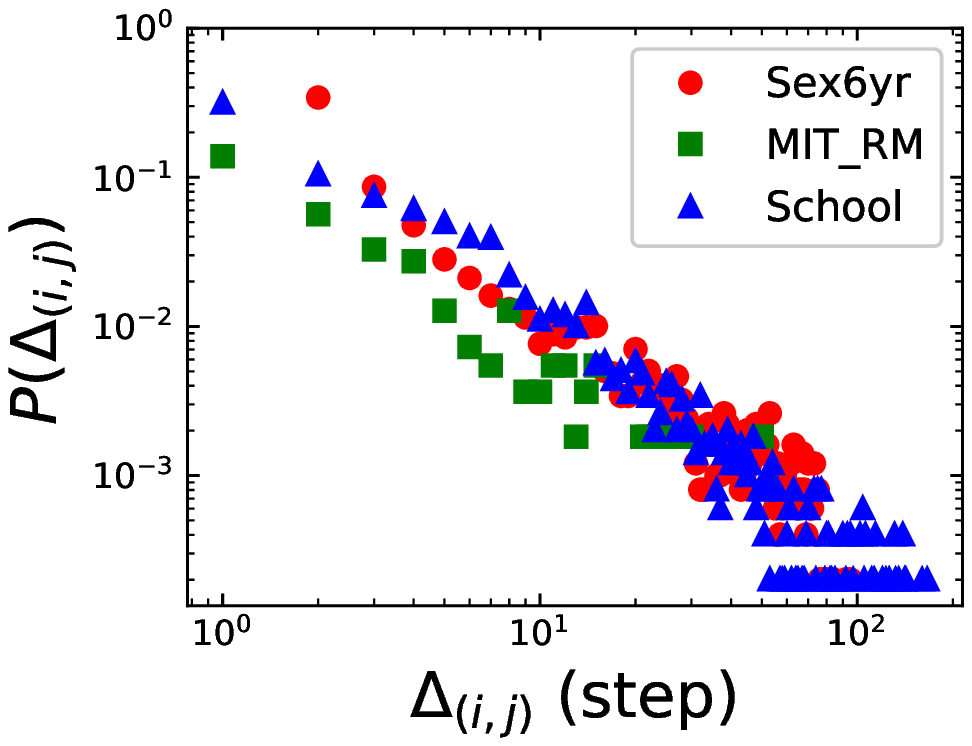}
\subcaption{}
\end{subfigure}
\begin{subfigure}[t]{0.3\textwidth}
\centering
\setlength{\abovecaptionskip}{0.cm}
\setlength{\belowcaptionskip}{-0.cm}
\includegraphics[width=1\textwidth]{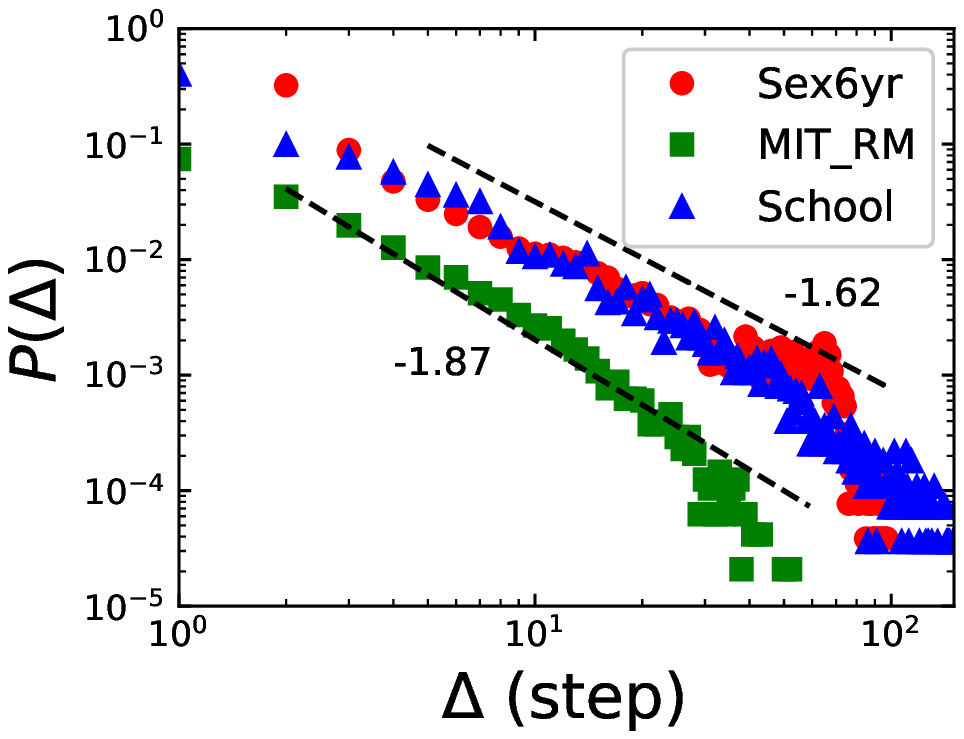}
\subcaption{}
\end{subfigure}
\caption{(a) Average fraction of social keeping contacts of each individual at each time step, with $p=19\%$ (in Sex6yr), $p=92\%$ (in MIR\_RM) and $p=92\%$ (in School). (b) Probability distributions $P(\Delta_{(i,j)})$ of the recurrence intervals $\Delta_{(i,j)}$ of the contacts $(i,j)$ for each individual $i$. Here we plot the distribution for 10 individuals as an example. (c) The inter-contact time distributions $P(\Delta)$ in empirical datasets, where $\Delta$ is the recurrence interval for any contact.}
\label{tieRetained}
\end{figure}

An active individual either establishes a contact by social keeping, connecting to a node which has already been connected to at previous time steps or social exploring, connecting to a new one. The average fraction of social keeping contacts of each individual at each time step is
\begin{equation}
p=\frac{1}{NT}\sum_{i=1}^N\sum_{t=1}^Tp_i(t),
\end{equation}
in which the fraction $p_i(t)$ of reconnected links of node $i$ is
\begin{equation}
p_i(t)=\frac{\sum_{j\in\mathcal{V}_{i}(t)}\delta_{ij}(t)}{|\mathcal{V}_{i}(t)|},
\end{equation}
where $\delta_{ij}(t)=0$ if agent $i$ never interacted with agent $j$ before, otherwise, $\delta_{ij}(t)=1$. The $|\mathcal{V}_{i}(t)|$ is the number of direct neighbors of agent $i$ at time step $t$. The fraction $p_i(t)$ quantifies the inclination of an agent to keep previous established contacts. Fig. \ref{tieRetained}(a) shows that in the context of sex trade, individuals dominantly adopt the strategy of social exploring, while, in the context of physical proximity (MIT\_RM) and face-to-face communication (School), individuals tend to reconnect with the nodes who have already contacted by social keeping.

Moreover, the recurrence interval of human contacts is analysed. The recurrence interval $\Delta_{(i,j)}$ of individuals $i$ and $j$ refers to the number of time steps between any two consecutive contacts. We can deduce that, if an active node $i$ has a degree $k_i$ at each time step and $k_i$ contacts are established randomly with other nodes, the distribution of recurrence interval is an exponential distribution as $P(\Delta_{(i,j)})=(k_i/N)(1-k_i/N)^{(\Delta_{(i,j)}-1)}$.  Note that this is the basic assumption of most previous works \cite{ubaldi2016asymptotic,ubaldi2017burstiness,karsai2014time,kim2015scaling,valdano2015predicting}. However, we find that the probability distributions of recurrence intervals of contacts between node $i$ and all others follow a power-law distribution in three datasets (see Fig. \ref{tieRetained}(b)). The found might give a support for the burst property of temporal networks \cite{barabasi2005origin}, that the inter-contact time distribution has a power-law form (see Fig. \ref{tieRetained}(c)). The results reply that agents do not establish contacts randomly with others regardless of their pervious contacts. Besides, we have identified the statistical law in real-world systems that agents preferentially contact individuals who have recently been in interact with \cite{jing2018quantifying}.

To summarize, two typical characteristics coexist in the process of realistic network evolution, \emph{individual activity state transition} and \emph{contact establishment}. In the next section we propose a temporal network model based on the mechanisms.
\section*{Modelling Dynamic Contact Networks}
\subsection*{Memory Driven (MD) Model}\label{MDmodel}
\begin{figure}[t]
\setlength{\abovecaptionskip}{0.cm}
\setlength{\belowcaptionskip}{-0.cm}
\centering
\includegraphics[width=3.5in]{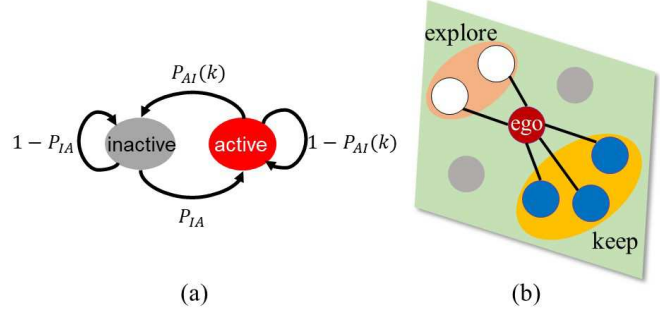}
\caption{(a) Transition probabilities of individual activity states in the MD network model. (b) Contact establishment strategies of each agent in each time step: social keeping (blue) and social exploring (white).}
\label{model}
\end{figure}

In this section, we present a temporal network model, named the memory driven (MD) model, of human contact networks. The empirical analysis of individual activity state transition and contact establishment are both the basic mechanisms for the MD model. We consider a set of agents $\mathcal{N}$ in a human contact network $\mathcal{G}$. For each time step, every agent has two possible activity states: active and inactive. We here assume that each individual has the maximum memory length $L$ time steps, in other words, the contact establishment at current time step is only influenced by the connections at pervious $L$ time steps, which is stored in the memory train $G_{\mathcal{M}}= \{G_{t-L+1}, G_{t-L+2}, ..., G_{t-1}, G_{t}\}$. We generate $L$ random networks \cite{erdos1960evolution} to initialize the temporal network $\mathcal{G} = \{G_1, G_2, ..., G_L\}$ and the memory train $G_{\mathcal{M}}=\{G_1, G_2, ..., G_L\}$. The generation of temporal network $\mathcal{G} = \{G_1, G_2, ..., G_T\}$ is illustrated in Fig. \ref{model} and described as follows:
\begin{enumerate}
  \item \textbf{Individual activity state transition}: an active individual $i$ with degree $k$ at time step $t$ becomes inactive at time step $t+1$ with probability $P_{AI}(k)=Ak^{-\alpha}$ ($A\in(0,1]$), while an inactive agent becomes active with a constant firing rate $P_{IA}$.
  \item \textbf{Contact establishment of active individuals}: We assign each active individual with $s$ contacts, where $s$ is extracted from a given probability distribution $F(s)$. The contacts are established by the following steps:
      \begin{enumerate}
        \item Social keeping: with a \emph{keeping rate} $p\in[0,1]$, agent $i$ connects to a previous contacted agent $j$ with probability $p_{ij}=f(\Delta_{(i,j)})\propto \Delta_{(i,j)}^{-\gamma},(1\leq\Delta_{(i,j)}\leq L)$, which is a function of the time interval $\Delta_{(i,j)}$ since their last interaction.
        \item Social exploring: with an \emph{exploring rate} $q=1-p$, the agent randomly interacts with a new agent who has never been contacted by agent $i$ before (or the previous contacts have exceeded the memory length $L$).
      \end{enumerate}
  \item Generate the network $G_{t+1}$ and update the memory train as $G_{\mathcal{M}}= \{G_{t-L+2}, G_{t-L+3}, ..., G_{t}, G_{t+1}\}$.
  \item Repeat (i)-(iii) until the end of time span $T$ of the temporal network.
\end{enumerate}

\begin{figure}[!ht]
\setlength{\abovecaptionskip}{0.cm}
\setlength{\belowcaptionskip}{-0.cm}
\centering
\begin{subfigure}[t]{0.35\textwidth}
\centering
\setlength{\abovecaptionskip}{0.cm}
\setlength{\belowcaptionskip}{-0.cm}
\includegraphics[width=1\textwidth]{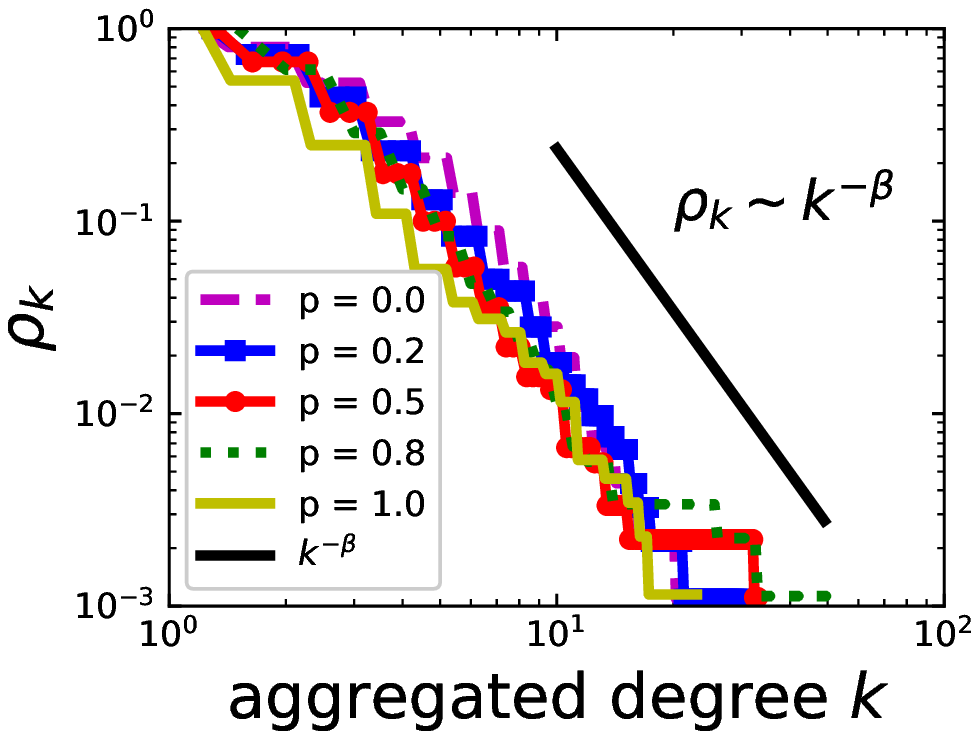}
\subcaption{}
\end{subfigure}
\begin{subfigure}[t]{0.35\textwidth}
\centering
\setlength{\abovecaptionskip}{0.cm}
\setlength{\belowcaptionskip}{-0.cm}
\includegraphics[width=1\textwidth]{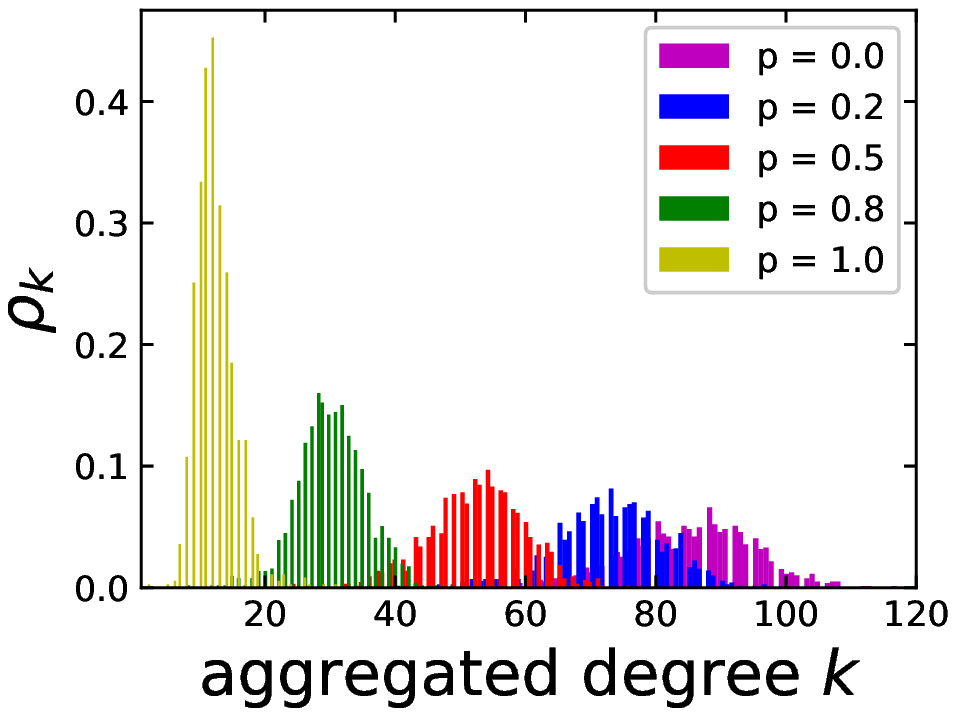}
\subcaption{}
\end{subfigure}
\begin{subfigure}[t]{0.35\textwidth}
\centering
\setlength{\abovecaptionskip}{0.cm}
\setlength{\belowcaptionskip}{-0.cm}
\includegraphics[width=1\textwidth]{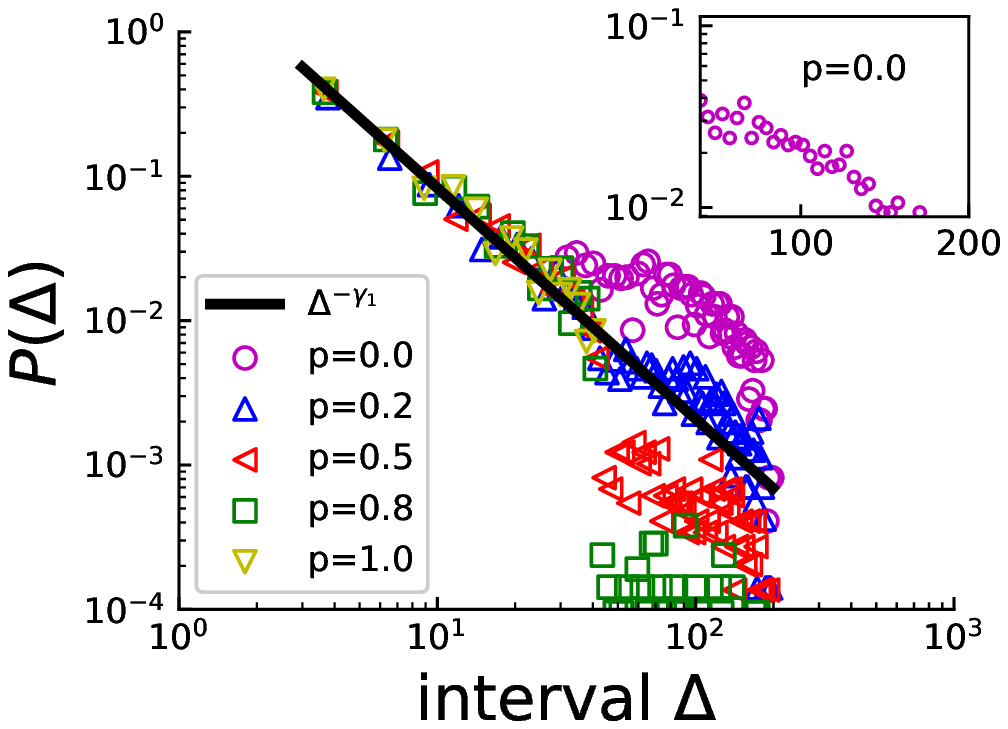}
\subcaption{}
\end{subfigure}
\begin{subfigure}[t]{0.35\textwidth}
\centering
\setlength{\abovecaptionskip}{0.cm}
\setlength{\belowcaptionskip}{-0.cm}
\includegraphics[width=1\textwidth]{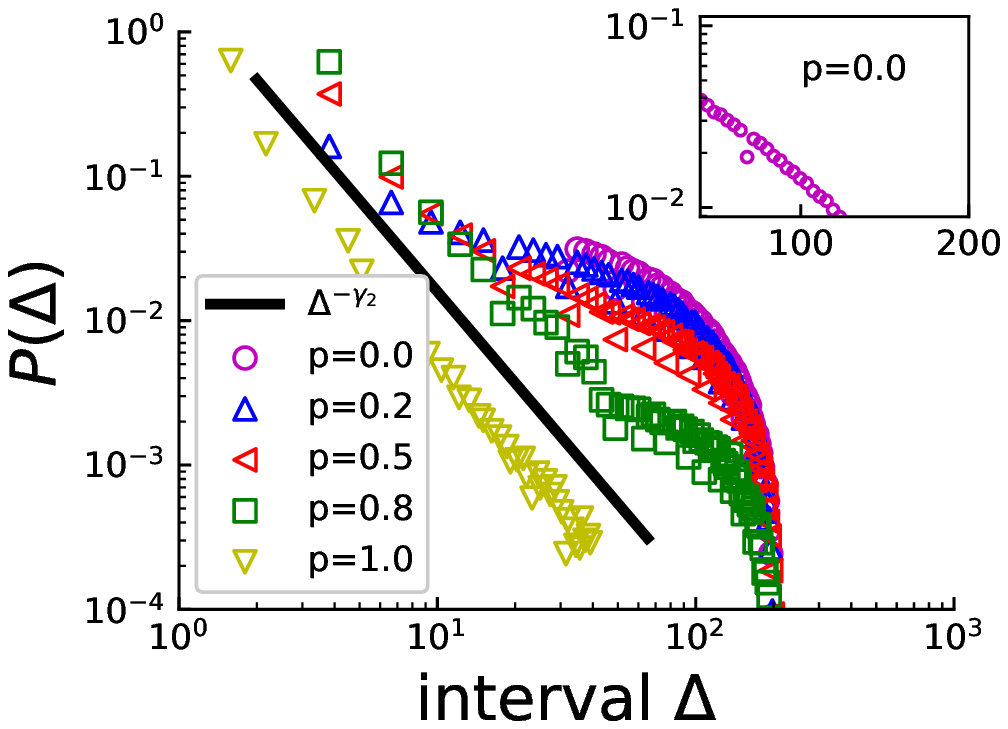}
\subcaption{}
\end{subfigure}
\caption{(a) (b) The degree distribution $\rho_k$ of integrated networks and (c) (d) inter-contact time distributions $P(\Delta)$ versus the keeping rate $p$ for one simulation. We fix $N=1000$, $A=1$, $\alpha=1$, $P_{IA}=0.1$, $L=40$ and $T=200$. In (a) and (c), the number of contacts $s$ is sampled from a power-law distribution $F(s)=(s/s_{min})^{-\beta}$ with $\beta=2.8$, $s_{min}=1$, and $\gamma_1=1.6$. In (b) and (d), the distribution $F(s)$ satisfies a Gaussian distribution $Norm(\mu,\sigma^2)$, with the mean $\mu=5$ and variance $\sigma^2=1$, $\gamma_2=1.8$. The inter-contact time follows an exponent distribution when $p=0$ (insets of (c) and (d)).}
\label{modelDIET}
\end{figure}

\subsection*{Analysis of the model}
Based on the above descriptions, we perform numerical simulations with two different parameter settings, that is, we assume that $F(s)$  follows a power-law distribution and a Gaussian distribution, respectively, to characterize the heterogeneity or homogeneity of human contact networks. The results in Figs. \ref{modelDIET}(a) and \ref{modelDIET}(b) show that the degree distributions $\rho_k$ of integrated networks generated by our model might be various for different $F(s)$. If $F(s)$ follows a power-law distribution, $\rho_k$ will have a power-law tail with the same exponent $\gamma$ regardless of the keeping rates $p$ (see Fig. \ref{modelDIET}(a)). If $F(s)$ follows a Gaussian distribution, the functional form of the degree distribution $\rho_k$ does not change with $p$, however, the average degree of integrated network decreases with the increase of $p$ (see Fig. \ref{modelDIET}(b)). The inter-contact time distributions $P(\Delta)$ of the two networks are shown in Figs. \ref{modelDIET}(c) and \ref{modelDIET}(d). When the keeping rate $p=0$, the establishment of human contacts is random and memoryless, which leads to an exponential distributions $P(\Delta)$ of the inter-contact time (insets of Figs. \ref{modelDIET}(c) and \ref{modelDIET}(d)). With the increase of $p$, the inter-contact time distribution asymptotically follows a power-law, indicating that the memory embedded in the process of contact establishment can induce the bursty interactive pattern of human activities.

\section*{Epidemic Processes in Temporal Human Contact Networks and Models}
\subsection*{Dynamical processes on real-world networks and network models}\label{models}
We study the Susceptible-Infected (SI) spreading dynamics \cite{kermack1927contribution,anderson1992infectious} on temporal human contact networks. The SI epidemic model, despite of its simplicity, has been pervasively operated as a useful tool to probe the topological structures and temporal characteristics of temporal networks \cite{vestergaard2014memory,karsai2011small,Starnini2013Immunization}. In this model, each agent can be in two possible states: susceptible (S) or infected (I). A susceptible (S) agent is infected by an infected (I) one with probability $\lambda$ if there is a contact between them. Once an agent is infected (I), it will not recover. We start each simulation with all agents in the susceptible state, and randomly select a single agent as the infected ``seed''. In order to uncover the effects of the two mechanisms, \emph{i.e.} individual activity state transition and contact establishment, on the network topologies, and further impacts on spreading dynamic processes on temporal networks, we proceed the SI spreading processes on real-world networks and four corresponding network models. The four network models include our memory driven (MD) model, 
 two null models (\emph{i.e.} AR model and CR model), and the activity driven (AD) model.

We here first introduce two null models where each of the mechanisms is separately destroyed. Besides, as a contrast, we reproduce the temporal network of empirical datasets with the activity driven model \cite{perra2012activity}, which has been widely used as a paradigmatic network example for the study of spreading dynamics taking place on the same time-scale of network evolution. The models are described in detail as follows:
\begin{enumerate}
  \item \textbf{Agent randomized (AR) null model}

  The AR null model only keep the contact establishment mechanism, that the recurrence interval of human contacts is preserved. We keep all the human interaction time in the empirical datasets unchanged, and replace each contact pair with two randomly selected individuals. Note that once a contact pair $(u_{1},v_{1})$ is replaced by $(u_{2},v_{2})$, all the contacts between individuals $u_{1}$ and $v_{1}$ will be replaced by contact pair $u_{2}$ and $v_{2}$.

  \item \textbf{Contact randomized (CR) null model}

  The CR null model only keep the individual activity state mechanism, that the degree and activity state of individuals are unchanged in each $G_{t}$. At each time step $t$, we randomly select two contacts associated with the four individuals, and then rewire the two contacts. If the two contacts both have the same individual, we discard the contact pair and randomly select two new contacts. The rewiring steps are repeated more than $2L_{C}(t)$ times to ensure the rewiring of most contacts in $G_{t}$, where $L_{C}(t)$ is the number of contacts in $G_{t}$. After the rewiring procedure, the interaction time and recurrence interval of human contacts are completely different from that of original interaction datasets.

  \item \textbf{Activity driven (AD) model}

  The AD model \cite{perra2012activity} is one of the most studied temporal network models. The generation of an AD model follows the rules: At each time step $t$, the $G_{t}$ starts with $N$ isolated nodes, and each node $i$ is assigned an activity probability $a_it_{w}$ to become active; Then, the active nodes generates $m$ links that are randomly connected to $m$ other nodes, and the inactive nodes can only receive connections from active nodes; At next time step $t+1$, all connections in $G_{t}$ are omitted, and the steps are repeated. Here we apply the AD model to generate temporal networks to match the real-world networks. The activity probability $a_i = \eta x_i$, where $x_i\in[\epsilon,1]$, $\epsilon$ is a lower cut-off that avoids possible divergences, and the $x_i$ is drawn from a given probability function $F(x_i)$. The probability function $F(x_i)$ is statistically obtained from empirical data, and the average number of active agents per unit time is $\eta\langle x\rangle N$, where $\eta=1$ in this work. The AD model does not capture the above two mechanisms in empirical data.
\end{enumerate}

\begin{figure}[t]
\centering
\includegraphics[width=1.0\textwidth]{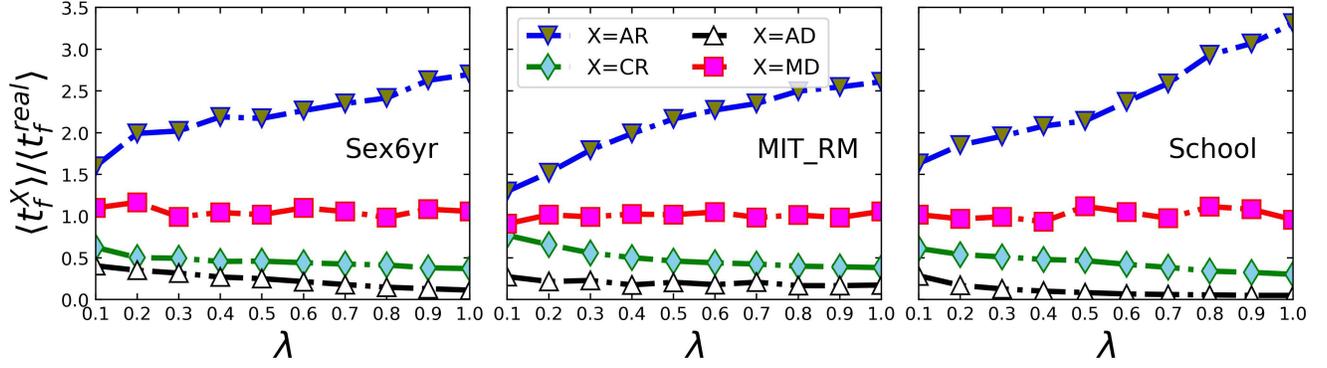}
\caption{The ratio $\langle t_f^X \rangle / \langle t_f^{real} \rangle$ of the average full prevalence time as a function of the infection rate $\lambda$ in three empirical datasets, where $X=AR, CR, AD, MD$ represent the agents randomized null models ($\bigtriangledown$), contacts randomized null models ($\diamondsuit$), activity driven models ($\bigtriangleup$) and memory driven models ($\Box$), respectively. The simulations are performed on $10^3$ realizations.}
\label{tf}
\end{figure}

\renewcommand{\arraystretch}{1.5}
\begin{table}[t]
\centering
\setlength{\abovecaptionskip}{0.cm}
\setlength{\belowcaptionskip}{-0.cm}
\caption{The parameters of memory driven model. $X\sim Pl(\beta)$ represents the power law $\rho(x)=(\beta-1)x^{-\beta}$, where $x\in[1,+\infty)$, and $X\sim Norm(\mu,\sigma^2)$ represents the Gaussian distribution with the mean $\mu$ and variance $\sigma^2$.}
\label{MDparameter}
\begin{tabular*}{0.85\textwidth}{@{\extracolsep{\fill}}c c c c}
\hline
\multicolumn{4}{c}{Memory Driven (MD) Model}\\
\hline
Parameters & Sex6yr & MIT\_RM & School \\
\hline
$|\mathcal{N}|$  & 16,706 & 96   &  236 \\
$T$      & 64     & 230  &  104 \\
$A$      & 0.78   & 0.45 &  0.33 \\
$\alpha$ & 0.84    & 0.56  &  0.91 \\
$P_{IA}$      & 0.04   & 0.16 &  0.27 \\
$F(s)$   & $Pl(-2.6)$ & $Norm(9.0,5.1)$ & $Norm(7.0,3.4)$ \\
$\gamma$ & 1.62   & 1.87 &  1.62 \\
$L$      & 36     & 60   &  12 \\
\hline
\end{tabular*}
\end{table}
\renewcommand{\arraystretch}{1.5}
\begin{table}[t]
\centering
\setlength{\abovecaptionskip}{0.cm}
\setlength{\belowcaptionskip}{-0.cm}
\caption{The parameters of activity driven model. $X\sim U(a,b)$ represents the uniform distribution from $a$ to $b$.}
\label{ADparameter}
\begin{tabular*}{0.82\textwidth}{@{\extracolsep{\fill}}c c c c}
\hline
\multicolumn{4}{c}{Activity Driven (AD) Model}\\
\hline
Parameters & Sex6yr & MIT\_RM & School \\
\hline
$|\mathcal{N}|$  & 16,706 & 96   &  236 \\
$T$    & 64      & 230  & 104 \\
$F(x)$ & $Pl(-2.6)$ & $U(0,1)$  & $U(0,1)$ \\
$\epsilon$ & $10^{-2}$   & $10^{-2}$ & $10^{-2}$ \\
$m$    & 1 & 5 & 3 \\
\hline
\end{tabular*}
\end{table}

In this work, the virus spreads on real human contact networks under a temporally periodic boundary condition (\emph{i.e.} repeating the whole contact sequence) and network models until all the reachable agents are infected \cite{karsai2011small}. The four models have been introduced in detail in Sections \ref{MDmodel} and \ref{models}. The relevant parameters of MD model and AD model are measured in the corresponding datasets (see details on Tables \ref{MDparameter} and \ref{ADparameter}). We repeat the SI spreading process for all possible seeds, and record the full prevalence time $t_f$, \emph{i.e.} the time to reach 100\% infection in a connected network or the largest connected component (LCC) of a disconnected network.

We calculate the ratio $\langle t_f^X(\lambda) \rangle / \langle t_f^{real}(\lambda) \rangle$ as a function of the infection rate $\lambda$, where $\langle t_f^{real}\rangle$ and $\langle t_f^X\rangle$ represent the average spreading time of a full prevalence in the real-world network $real$ and the temporal network model $X$, respectively. If the ratio $\langle t_f^X(\lambda) \rangle / \langle t_f^{real}(\lambda) \rangle=1$, the full prevalence time of a spreading in the real-world network is the same as that in the corresponding network model $X$. In other words, the $X$ is a reasonable temporal model for human contact networks in the study of spreading processes. As shown in Fig. \ref{tf}, the results in three empirical datasets are similar. The ratio $\langle t_f^{AR}\rangle/\langle t_f^{real}\rangle$ is always larger than $1$ regardless of the infection rate $\lambda$, and the ratio $\langle t_f^{AR}\rangle/\langle t_f^{real}\rangle $ increases with the increase of the infection rate $\lambda$. The results illustrate that the individual state transition mechanism might promote the spreading processes, since the spread of virus in the AR model, which only keeps the contact establishment mechanism and omits the individual activity state transition mechanism, is faster than that in real human contact networks. Moreover, the promotion of spreading is even larger when the infection rate $\lambda$ is lager. However, there is a very small difference between the CR model and AD model, where the ratios $\langle t_f^{CR}\rangle/\langle t_f^{real}\rangle$ and $\langle t_f^{AD}\rangle/\langle t_f^{real}\rangle$ are both much smaller than $1$. Both models are built without the contact establishment mechanism, which might cause a slowing-down effect on the spreading dynamics. We find that the ratio $\langle t_f^{MD}\rangle/\langle t_f^{real}\rangle$ always fluctuates around $1$, and remains stable against the infection rate $\lambda$. Fig. \ref{tf} confirms that the MD model is superior to all other models in characterizing human contact networks. This phenomenon highlights the crucial role of the two mechanisms considered in our model. They both serve as an indispensable component to characterize the evolution of the real-world networks. Binding them together enables us to accurately capture the contagion processes unfolding on empirical human contact networks.


\subsection*{Dynamical processes on memory-driven networks}
\begin{figure}[t]
\centering
\includegraphics[width=0.45\textwidth]{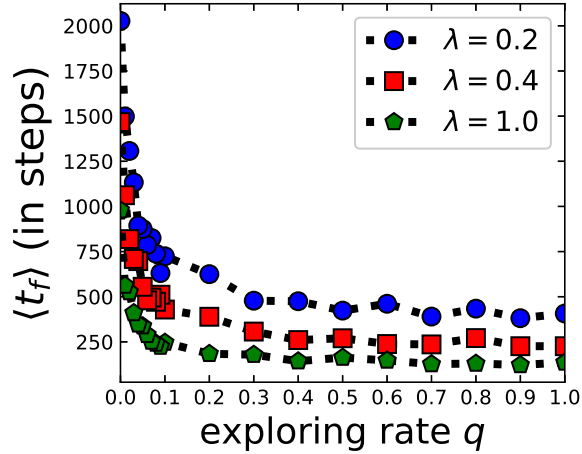}
\caption{The average full prevalence time as a function of the exploring rate $q$ versus the infection rate $\lambda$. The temporal networks are generated by the MD model with $N=1000$, $A=1$, $\alpha=1$, $\gamma=1.6$, $P_{IA}=0.1$, $L=40$, $T=3000$, and $s$ is sampled from $F(s)=(s/s_{min})^{-\beta}$ with $\beta=2.8$, $s_{min}=1$. The simulation result is the average of 200 realizations.}
\label{varyP}
\end{figure}

 We here study how the network evolution affects dynamic processes, such as the epidemic spreading on networks. The exploring rate $q$ in the contact establishment mechanism of MD model can tune the evolution of network topology. We perform the SI spreading processes with infection rates $\lambda=0.2, 0.4$ and 1.0 on temporal networks generated by the MD model with $N = 1000$ and the exploring rates $q=0, 0.1,...,1$, respectively, and record the full prevalence time $t_f$.
Fig. \ref{varyP} indicates that the $t_f$ is extremely large when $q = 0$, which implies that all contacts in current time step are all recurrences of previously established contacts, that is, no individual is willing to explore a new social relationship in the evolving network. Moreover, we observe that the average full prevalence time $\langle t_f\rangle$ declines rapidly with the exploring rate $q$, when $q$ is small. When the exploring rate $q$ further increases and reaches a specific value (around 0.1), the spreading time decreases slightly and finally meets a saturation. The fact illustrates that even a small percentage of individuals would like to explore new social ties, the spreading process will be promoted, however, when $q$ is larger than a specific value, the influence of $q$ on the spreading process is rather limited.

Furthermore, the SI spreading model can be used to characterize the information spreading on human contact networks. The temporal property, \emph{i.e.} the exploring rate, could speed up the spreading processes occurring on networks intentionally. In real-world social systems, the establishment of social ties is costly, and the cost of exploring new social relationships is higher than that of maintaining old social ties \cite{dunbar2009social,tamarit2018cognitive,dunbar2018social}. In order to increase the speed of information dissemination and control the social cost at the meanwhile, according to our findings, encouraging a small percentage of individuals to establish new social ties could lead to an explosive propagation in networks. The finding can be applied to many scenarios, such as promoting the product marketing in social networks and accelerating the information diffusion on social media.

\section*{Conclusion}
We have presented the analysis of several empirical datasets of human contacts. We observe two crucial mechanisms, \emph{i.e.} the individual activity state transition and contact establishment, that create the evolving structures of human contact networks. The individual activity state transition governs the activation of agents. The active agents establish contacts with others according to the strategies of contact establishment, that is, agents tend to interact with recently contacted one. We find that the transition probability $P_{AI}(k)$ or $P_{AA}(k)$ between the active state and the inactive state is a power-law function of the degree $k$. Moreover, the empirical datasets show that not all active nodes have the number of contacts at each time step, and the recurrence intervals of individual contact establishment follow a power-law distribution. Considering the empirical observations, we propose a novel temporal network model (the memory driven model) based on the two mechanisms. Furthermore, we study the effects of the two mechanisms on dynamical processes. We perform the SI spreading on the real-world human contact networks and four temporal network models. The full prevalence times of SI spreading on the networks and models are compared. Our model shows a good agreement with three empirical temporal networks, which implies that the two mechanisms enable us to capture the evolution of human interactions in temporal networks. Interestingly, the results demonstrate that the individual activity state transition accelerates the diffusion processes, contrarily, the specific contact establishment strategy slows down the spreading. Besides, we find that the exploration of new social ties effectively promote the spreading processes, and a small percentage of individual exploring new social ties is sufficient to induce an explosive spreading on networks. The study thus paves the way to a better understanding of the mechanisms driving the evolution of human contacts and their effects on dynamic processes in real-world social systems.
\section*{Acknowledgments}
This work was supported in part by the National Natural Science Foundation of China (No. 71731004, No.61603097), in part by the Natural Science Fund for distinguished Young Scholar of China (No. 61425019), and Natural Science Foundation of Shanghai (No.16ZR1446400).

\section*{Appendix}
\subsection*{A. Data Description}
\begin{enumerate}
  \item The dataset of ``Sex6yr" was gathered from an online forum in Brazil from September 2002 to October 2008. Note that although the data belongs to the online survey, each record reflects a real offline sex trading activity among the sellers and buyers. We use the dataset after discarding the initial transient (312 days) of community growth \cite{Rocha2010Information}. The dataset is available in \cite{rocha2011simulated}, and more details are described in \cite{Rocha2010Information}.
  \item The dataset of ``MIT\_RM" was recorded by the Reality Mining Project conducted by MIT Media Lab from September 2004 to May 2005. In the project, the subjects were required to use mobile phones with pre-installed softwares, which could sample their physical proximity via bluetooth devices every 6 minutes and record the corresponding user tags. The dataset is available on the website of Reality Commons (http://realitycommons.media.mit.edu). From the dataset we have removed 2 days as they are empty of human activities. We refer to \cite{Nathan2006Reality} for more details on the data description and collection strategy.
  \item The dataset of ``School" was collected in a French school on October 1, 2009 by the SocioPatterns collaboration. The data recorded the time-resolved face-to-face proximity of children and teachers, with Radio-Frequency IDentification (RFID) device embedded in badges. The dataset is available on the website of SocioPatterns (http://www.sociopattern.org/datasets), and more details are introduced in \cite{Stehl2011High}.
\end{enumerate}

\subsection*{B. Selection of Time Window}
The selection of time window size is crucial to analyze the evolution of the network structure.
If $t_w$ is too fine, the temporal contacts are aggregated over insufficient time. The resulting network is too sparse and messy, which makes it difficult to observe interesting phenomena such as the formation of a giant component or the disappearance of a cluster \cite{sekara2016fundamental}. Conversely, if $t_w$ is too coarse, the aggregated network will not be able to capture the critical temporal information such as link concurrency, time-respecting path and reachability \cite{Holme2012Temporal}.
If $t_w$ is large enough to aggregate all the contacts observed into a single time slice, the temporal network is degraded into a static network.
Therefore, an appropriate time window should strike a balance between the disturbation that disguise the relevant topological changes (small $t_w$) and the loss of temporal structural information (large $t_w$).

Inspired by the method introduced in \cite{sekara2016fundamental,sulo2010meaningful,caceres2011temporal}, here we consider the correlation between adjacent networks. First, we segment the empirical data into adjacency time steps of length $t_{w}=w\triangle t$, where $\triangle t$ is the resolution, the $w$ is the number of resolutions in a time step. The contacts within the time interval $[(t-1)t_w, t\cdot t_w)$ is aggregated into a static undirected graph $G_t$. Once $t_w$ is set, the temporal network is represented as a sequence of networks in time order $\mathcal{G}(t_w) = \{G_1, G_2, ..., G_t, ..., G_T\}$, where $T$ is the total number of time steps. The network $G_t= (\mathcal{N}_t, \mathcal{L}_t)$ at time step $t$ consists of a set of nodes $\mathcal{N}_t$ connected by a set of links $\mathcal{L}_t$. Then, by the Jaccard index \cite{sekara2016fundamental}, we calculate the link overlap $\sigma_L$ between adjacent networks $G_{t-1}= (\mathcal{N}_{t-1}, \mathcal{L}_{t-1})$ and $G_t= (\mathcal{N}_t, \mathcal{L}_t)$ as
\begin{equation}
\sigma_L(t,t_w)=\frac{|\mathcal{L}_{t-1}\cap\mathcal{L}_{t}|}{|\mathcal{L}_{t-1}\cup\mathcal{L}_{t}|}.
\end{equation}
The $\sigma_L$ takes values from the interval $[0, 1]$, with $\sigma_L=0$ indicating that the adjacent networks share no common link, and $\sigma_L=1$ indicating that the same network is exactly retained, \emph{i.e.} $G_t = G_{t-1}$. The average correlation across all adjacent networks can be calculated with
\begin{equation}
\sigma_L(t_w)=\frac{1}{T}\sum_{t=1}^T\sigma_L(t,t_w).
\end{equation}
\begin{figure*}[!ht]
\centering
\begin{subfigure}[t]{0.31\textwidth}
\setlength{\abovecaptionskip}{0.cm}
\setlength{\belowcaptionskip}{-0.cm}
\centering
\includegraphics[width=1\textwidth]{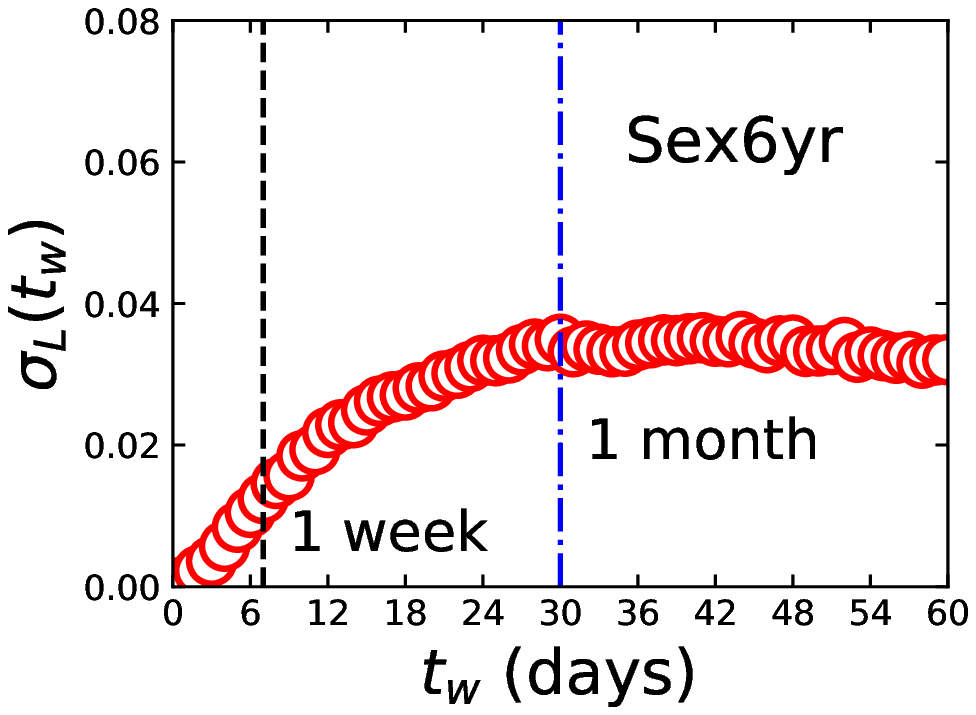}
\subcaption{}
\end{subfigure}
\begin{subfigure}[t]{0.31\textwidth}
\setlength{\abovecaptionskip}{0.cm}
\setlength{\belowcaptionskip}{-0.cm}
\centering
\includegraphics[width=1\textwidth]{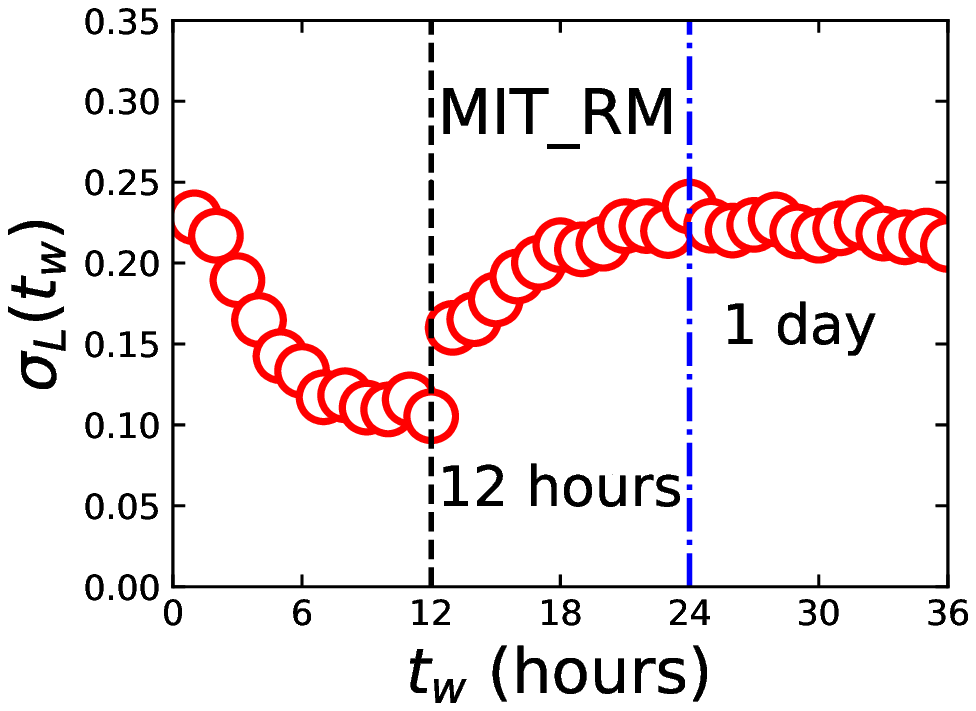}
\subcaption{}
\end{subfigure}
\begin{subfigure}[t]{0.31\textwidth}
\setlength{\abovecaptionskip}{0.cm}
\setlength{\belowcaptionskip}{-0.cm}
\centering
\includegraphics[width=1\textwidth]{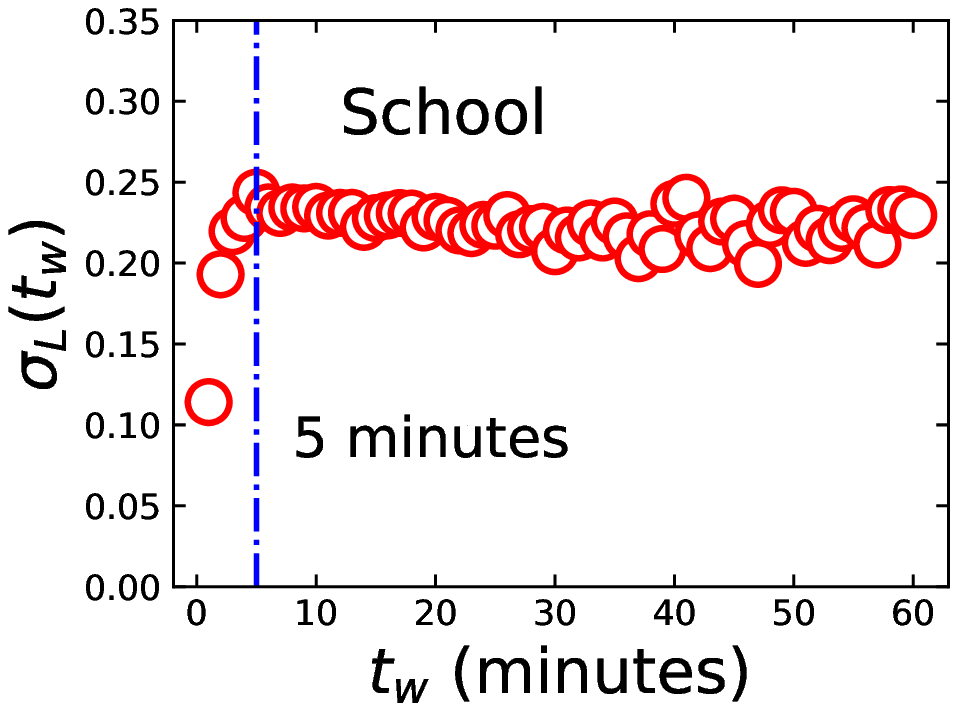}
\subcaption{}
\end{subfigure}
\caption{The average correlation as a function of the time window $t_w$. The values were averaged over all pairs of adjacent networks. }
\label{window}
\end{figure*}

As shown in Figs. \ref{window}(a)-(c), the average fraction of link overlap in each empirical dataset hits a peak when the time window $t_w$ is selected as 1 month (Sex6yr), 1 day (MIT\_RM) and 5 minutes (School), respectively. This phenomenon indicates that the adjacent networks are highly correlated under the corresponding time window, which characterizes the contact memory encoded in human interaction behaviors. As $t_w$ further increases, the average fraction of link overlap begins to slowly decline. This can be explained by our observation that agents with a low activity gradually make random memoryless contacts, and subsequently join into the aggregated network \cite{jing2018quantifying}. Therefore, according to the trade-off principle mentioned earlier, we aggregate the temporal contacts in empirical datasets with the time window corresponding to the peak point, and denote it as $w_0$. The time window size $w_0$ for three empirical datasets are selected as 1 month (Sex6yr), 1 day (MIT\_RM) and 5 minutes (School), respectively. 


\bibliographystyle{unsrt}
\bibliography{ref}

\begin{thebibliography}{10}
\expandafter\ifx\csname url\endcsname\relax
  \def\url#1{\texttt{#1}}\fi
\expandafter\ifx\csname urlprefix\endcsname\relax\def\urlprefix{URL }\fi
\providecommand{\bibinfo}[2]{#2}
\providecommand{\eprint}[2][]{\url{#2}}

\bibitem{granovetter1973strength}
\bibinfo{author}{Granovetter, M.~S.}
\newblock \bibinfo{title}{The strength of weak ties}.
\newblock \emph{\bibinfo{journal}{American journal of sociology}}
  \textbf{\bibinfo{volume}{78}}, \bibinfo{pages}{1360--1380}
  (\bibinfo{year}{1973}).

\bibitem{krackhardt2003strength}
\bibinfo{author}{Krackhardt, D.}, \bibinfo{author}{Nohria, N.} \&
  \bibinfo{author}{Eccles, B.}
\newblock \bibinfo{title}{The strength of strong ties}.
\newblock \emph{\bibinfo{journal}{Networks in the knowledge economy}}
  \bibinfo{pages}{82} (\bibinfo{year}{2003}).

\bibitem{starnini2013modeling}
\bibinfo{author}{Starnini, M.}, \bibinfo{author}{Baronchelli, A.} \&
  \bibinfo{author}{Pastor-Satorras, R.}
\newblock \bibinfo{title}{Modeling human dynamics of face-to-face interaction
  networks}.
\newblock \emph{\bibinfo{journal}{Phys. Rev. Lett.}}
  \textbf{\bibinfo{volume}{110}}, \bibinfo{pages}{168701}
  (\bibinfo{year}{2013}).

\bibitem{Zhang2015Human}
\bibinfo{author}{Zhang, Y.-Q.}, \bibinfo{author}{Li, X.}, \bibinfo{author}{Xu,
  J.} \& \bibinfo{author}{Vasilakos, A.}
\newblock \bibinfo{title}{Human interactive patterns in temporal networks}.
\newblock \emph{\bibinfo{journal}{IEEE Trans. Syst.,Man, Cybern. A, Syst.,
  Humans}} \textbf{\bibinfo{volume}{45}}, \bibinfo{pages}{214--222}
  (\bibinfo{year}{2015}).

\bibitem{zhang2015characterizing}
\bibinfo{author}{Zhang, Y.-Q.}, \bibinfo{author}{Li, X.},
  \bibinfo{author}{Liang, D.} \& \bibinfo{author}{Cui, J.}
\newblock \bibinfo{title}{Characterizing bursts of aggregate pairs with
  individual poissonian activity and preferential mobility}.
\newblock \emph{\bibinfo{journal}{IEEE Commun. Lett.}}
  \textbf{\bibinfo{volume}{19}}, \bibinfo{pages}{1225--1228}
  (\bibinfo{year}{2015}).

\bibitem{Nathan2006Reality}
\bibinfo{author}{Nathan~Eagle, A.~P.}
\newblock \bibinfo{title}{Reality mining: Sensing complex social systems}.
\newblock In \emph{\bibinfo{booktitle}{J. of Personal and Ubiquitous
  Computing}}, \bibinfo{pages}{255--268} (\bibinfo{year}{2006}).

\bibitem{dong2011modeling}
\bibinfo{author}{Dong, W.}, \bibinfo{author}{Lepri, B.} \&
  \bibinfo{author}{Pentland, A.~S.}
\newblock \bibinfo{title}{Modeling the co-evolution of behaviors and social
  relationships using mobile phone data}.
\newblock In \emph{\bibinfo{booktitle}{Proceedings of the 10th International
  Conference on Mobile and Ubiquitous Multimedia}}, \bibinfo{pages}{134--143}
  (\bibinfo{organization}{ACM}, \bibinfo{year}{2011}).

\bibitem{Cattuto2010Dynamics}
\bibinfo{author}{Cattuto, C.}, \bibinfo{author}{Broeck, W. V.~D.},
  \bibinfo{author}{Barrat, A.}, \bibinfo{author}{Colizza, V.} \&
  \bibinfo{author}{Vespignani, A.}
\newblock \bibinfo{title}{Dynamics of person-to-person interactions from
  distributed rfid sensor networks.}
\newblock \emph{\bibinfo{journal}{PLoS ONE}} \textbf{\bibinfo{volume}{5}},
  \bibinfo{pages}{e11596} (\bibinfo{year}{2010}).

\bibitem{isella2011s}
\bibinfo{author}{Isella, L.} \emph{et~al.}
\newblock \bibinfo{title}{What's in a crowd? {A}nalysis of face-to-face
  behavioral networks}.
\newblock \emph{\bibinfo{journal}{J. Theor. Biol.}}
  \textbf{\bibinfo{volume}{271}}, \bibinfo{pages}{166--180}
  (\bibinfo{year}{2011}).

\bibitem{saramaki2015seconds}
\bibinfo{author}{Saram{\"a}ki, J.} \& \bibinfo{author}{Moro, E.}
\newblock \bibinfo{title}{From seconds to months: an overview of multi-scale
  dynamics of mobile telephone calls}.
\newblock \emph{\bibinfo{journal}{Eur. Phys. J. B}}
  \textbf{\bibinfo{volume}{88}}, \bibinfo{pages}{164} (\bibinfo{year}{2015}).

\bibitem{saramaki2014persistence}
\bibinfo{author}{Saram{\"a}ki, J.} \emph{et~al.}
\newblock \bibinfo{title}{Persistence of social signatures in human
  communication}.
\newblock \emph{\bibinfo{journal}{Proc. Natl. Acad. Sci. USA}}
  \textbf{\bibinfo{volume}{111}}, \bibinfo{pages}{942--947}
  (\bibinfo{year}{2014}).

\bibitem{sekara2016fundamental}
\bibinfo{author}{Sekara, V.}, \bibinfo{author}{Stopczynski, A.} \&
  \bibinfo{author}{Lehmann, S.}
\newblock \bibinfo{title}{Fundamental structures of dynamic social networks}.
\newblock \emph{\bibinfo{journal}{Proc. Natl. Acad. Sci. USA}}
  \textbf{\bibinfo{volume}{113}}, \bibinfo{pages}{9977--9982}
  (\bibinfo{year}{2016}).

\bibitem{ubaldi2016asymptotic}
\bibinfo{author}{Ubaldi, E.} \emph{et~al.}
\newblock \bibinfo{title}{Asymptotic theory of time-varying social networks
  with heterogeneous activity and tie allocation}.
\newblock \emph{\bibinfo{journal}{Sci. Rep.}} \textbf{\bibinfo{volume}{6}},
  \bibinfo{pages}{35724} (\bibinfo{year}{2016}).

\bibitem{ubaldi2017burstiness}
\bibinfo{author}{Ubaldi, E.}, \bibinfo{author}{Vezzani, A.},
  \bibinfo{author}{Karsai, M.}, \bibinfo{author}{Perra, N.} \&
  \bibinfo{author}{Burioni, R.}
\newblock \bibinfo{title}{Burstiness and tie activation strategies in
  time-varying social networks}.
\newblock \emph{\bibinfo{journal}{Sci. Rep.}} \textbf{\bibinfo{volume}{7}},
  \bibinfo{pages}{46225} (\bibinfo{year}{2017}).

\bibitem{Medus2014Memory}
\bibinfo{author}{Medus, A.~D.} \& \bibinfo{author}{Dorso, C.~O.}
\newblock \bibinfo{title}{Memory effects induce structure in social networks
  with activity-driven agents}.
\newblock \emph{\bibinfo{journal}{J. Stat. Mech.}}
  \textbf{\bibinfo{volume}{2014}}, \bibinfo{pages}{09009}
  (\bibinfo{year}{2014}).

\bibitem{barabasi2005origin}
\bibinfo{author}{Barabasi, A.-L.}
\newblock \bibinfo{title}{The origin of bursts and heavy tails in human
  dynamics}.
\newblock \emph{\bibinfo{journal}{Nature}} \textbf{\bibinfo{volume}{435}},
  \bibinfo{pages}{207--211} (\bibinfo{year}{2005}).

\bibitem{karsai2011small}
\bibinfo{author}{Karsai, M.} \emph{et~al.}
\newblock \bibinfo{title}{Small but slow world: How network topology and
  burstiness slow down spreading}.
\newblock \emph{\bibinfo{journal}{Phys. Rev. E}} \textbf{\bibinfo{volume}{83}},
  \bibinfo{pages}{025102} (\bibinfo{year}{2011}).

\bibitem{Rocha2013Bursts}
\bibinfo{author}{Rocha, L.~E.} \& \bibinfo{author}{Blondel, V.~D.}
\newblock \bibinfo{title}{Bursts of vertex activation and epidemics in evolving
  networks}.
\newblock \emph{\bibinfo{journal}{PLoS Comput. Biol.}}
  \textbf{\bibinfo{volume}{9}}, \bibinfo{pages}{e1002974}
  (\bibinfo{year}{2013}).

\bibitem{miritello2013limited}
\bibinfo{author}{Miritello, G.}, \bibinfo{author}{Lara, R.},
  \bibinfo{author}{Cebrian, M.} \& \bibinfo{author}{Moro, E.}
\newblock \bibinfo{title}{Limited communication capacity unveils strategies for
  human interaction}.
\newblock \emph{\bibinfo{journal}{Sci. Rep.}} \textbf{\bibinfo{volume}{3}},
  \bibinfo{pages}{1950} (\bibinfo{year}{2013}).

\bibitem{vestergaard2014memory}
\bibinfo{author}{Vestergaard, C.~L.}, \bibinfo{author}{G{\'e}nois, M.} \&
  \bibinfo{author}{Barrat, A.}
\newblock \bibinfo{title}{How memory generates heterogeneous dynamics in
  temporal networks}.
\newblock \emph{\bibinfo{journal}{Phys. Rev. E}} \textbf{\bibinfo{volume}{90}},
  \bibinfo{pages}{42805} (\bibinfo{year}{2014}).

\bibitem{Scholtes2014Causality}
\bibinfo{author}{Scholtes, I.} \emph{et~al.}
\newblock \bibinfo{title}{Causality-driven slow-down and speed-up of diffusion
  in non-markovian temporal networks.}
\newblock \emph{\bibinfo{journal}{Nat. Commun.}} \textbf{\bibinfo{volume}{5}},
  \bibinfo{pages}{5024} (\bibinfo{year}{2014}).

\bibitem{perra2012activity}
\bibinfo{author}{Perra, N.}, \bibinfo{author}{Gon{\c{c}}alves, B.},
  \bibinfo{author}{Pastor-Satorras, R.} \& \bibinfo{author}{Vespignani, A.}
\newblock \bibinfo{title}{Activity driven modeling of time varying networks}.
\newblock \emph{\bibinfo{journal}{Sci. Rep.}} \textbf{\bibinfo{volume}{2}},
  \bibinfo{pages}{469} (\bibinfo{year}{2012}).

\bibitem{hoppe2013mutual}
\bibinfo{author}{Hoppe, K.} \& \bibinfo{author}{Rodgers, G.}
\newblock \bibinfo{title}{Mutual selection in time-varying networks}.
\newblock \emph{\bibinfo{journal}{Phys. Rev. E}} \textbf{\bibinfo{volume}{88}},
  \bibinfo{pages}{042804} (\bibinfo{year}{2013}).

\bibitem{pozzana2017epidemic}
\bibinfo{author}{Pozzana, I.}, \bibinfo{author}{Sun, K.} \&
  \bibinfo{author}{Perra, N.}
\newblock \bibinfo{title}{Epidemic spreading on activity-driven networks with
  attractiveness}.
\newblock \emph{\bibinfo{journal}{Phys. Rev. E}} \textbf{\bibinfo{volume}{96}},
  \bibinfo{pages}{042310} (\bibinfo{year}{2017}).

\bibitem{alessandretti2017random}
\bibinfo{author}{Alessandretti, L.}, \bibinfo{author}{Sun, K.},
  \bibinfo{author}{Baronchelli, A.} \& \bibinfo{author}{Perra, N.}
\newblock \bibinfo{title}{Random walks on activity-driven networks with
  attractiveness}.
\newblock \emph{\bibinfo{journal}{Phys. Rev. E}} \textbf{\bibinfo{volume}{95}},
  \bibinfo{pages}{052318} (\bibinfo{year}{2017}).

\bibitem{karsai2014time}
\bibinfo{author}{Karsai, M.}, \bibinfo{author}{Perra, N.} \&
  \bibinfo{author}{Vespignani, A.}
\newblock \bibinfo{title}{Time varying networks and the weakness of strong
  ties}.
\newblock \emph{\bibinfo{journal}{Sci. Rep.}} \textbf{\bibinfo{volume}{4}},
  \bibinfo{pages}{4001} (\bibinfo{year}{2014}).

\bibitem{kim2015scaling}
\bibinfo{author}{Kim, H.}, \bibinfo{author}{Ha, M.} \& \bibinfo{author}{Jeong,
  H.}
\newblock \bibinfo{title}{Scaling properties in time-varying networks with
  memory}.
\newblock \emph{\bibinfo{journal}{Eur. Phys. J. B}}
  \textbf{\bibinfo{volume}{88}}, \bibinfo{pages}{1--8} (\bibinfo{year}{2015}).

\bibitem{valdano2015predicting}
\bibinfo{author}{Valdano, E.} \emph{et~al.}
\newblock \bibinfo{title}{Predicting epidemic risk from past temporal contact
  data}.
\newblock \emph{\bibinfo{journal}{PLoS Comput. Biol.}}
  \textbf{\bibinfo{volume}{11}}, \bibinfo{pages}{e1004152}
  (\bibinfo{year}{2015}).

\bibitem{jing2018quantifying}
\bibinfo{author}{Li, J.}, \bibinfo{author}{Li, C.} \& \bibinfo{author}{Li, X.}
\newblock \bibinfo{title}{Quantifying the contact memory in temporal human
  interactions}.
\newblock In \emph{\bibinfo{booktitle}{2018 IEEE International Symposium on
  Circuits and Systems (ISCAS)}}, \bibinfo{pages}{1--5} (\bibinfo{year}{2018}).

\bibitem{barabasi1999emergence}
\bibinfo{author}{Barab{\'a}si, A.-L.} \& \bibinfo{author}{Albert, R.}
\newblock \bibinfo{title}{Emergence of scaling in random networks}.
\newblock \emph{\bibinfo{journal}{science}} \textbf{\bibinfo{volume}{286}},
  \bibinfo{pages}{509--512} (\bibinfo{year}{1999}).

\bibitem{Romualdo2001Epidemic}
\bibinfo{author}{Pastor-Satorras, R.} \& \bibinfo{author}{Vespignani, A.}
\newblock \bibinfo{title}{Epidemic spreading in scale-free networks}.
\newblock \emph{\bibinfo{journal}{Phys. Rev. Lett.}}
  \textbf{\bibinfo{volume}{86}}, \bibinfo{pages}{3200} (\bibinfo{year}{2001}).

\bibitem{Castellano2010Thresholds}
\bibinfo{author}{Castellano, C.} \& \bibinfo{author}{Pastorsatorras, R.}
\newblock \bibinfo{title}{Thresholds for epidemic spreading in networks.}
\newblock \emph{\bibinfo{journal}{Phys. Rev. Lett.}}
  \textbf{\bibinfo{volume}{105}}, \bibinfo{pages}{218701}
  (\bibinfo{year}{2010}).

\bibitem{Wang2016Identifying}
\bibinfo{author}{Wang, J.~B.}, \bibinfo{author}{Wang, L.} \&
  \bibinfo{author}{Li, X.}
\newblock \bibinfo{title}{Identifying spatial invasion of pandemics on
  metapopulation networks via anatomizing arrival history}.
\newblock \emph{\bibinfo{journal}{IEEE Trans. Cybern.}}
  \textbf{\bibinfo{volume}{46}}, \bibinfo{pages}{2782--2795}
  (\bibinfo{year}{2016}).

\bibitem{Qu2017Ranking}
\bibinfo{author}{Qu, B.}, \bibinfo{author}{Li, C.}, \bibinfo{author}{Mieghem,
  P.~V.} \& \bibinfo{author}{Wang, H.}
\newblock \bibinfo{title}{Ranking of nodal infection probability in
  susceptible-infected-susceptible epidemic}.
\newblock \emph{\bibinfo{journal}{Sci. Rep.}} \textbf{\bibinfo{volume}{7}},
  \bibinfo{pages}{9233} (\bibinfo{year}{2017}).

\bibitem{Iribarren2009Impact}
\bibinfo{author}{Iribarren, J.~L.} \& \bibinfo{author}{Moro, E.}
\newblock \bibinfo{title}{Impact of human activity patterns on the dynamics of
  information diffusion}.
\newblock \emph{\bibinfo{journal}{Phys. Rev. Lett.}}
  \textbf{\bibinfo{volume}{103}}, \bibinfo{pages}{038702}
  (\bibinfo{year}{2009}).

\bibitem{Montanari2010The}
\bibinfo{author}{Montanari, A.} \& \bibinfo{author}{Saberi, A.}
\newblock \bibinfo{title}{The spread of innovations in social networks.}
\newblock \emph{\bibinfo{journal}{Proc. Natl. Acad. Sci. USA}}
  \textbf{\bibinfo{volume}{107}}, \bibinfo{pages}{20196--201}
  (\bibinfo{year}{2010}).

\bibitem{kreindler2014rapid}
\bibinfo{author}{Kreindler, G.~E.} \& \bibinfo{author}{Young, H.~P.}
\newblock \bibinfo{title}{Rapid innovation diffusion in social networks}.
\newblock \emph{\bibinfo{journal}{Proc. Natl. Acad. Sci. USA}}
  \textbf{\bibinfo{volume}{111}}, \bibinfo{pages}{10881--10888}
  (\bibinfo{year}{2014}).

\bibitem{erdos1960evolution}
\bibinfo{author}{Erdos, P.} \& \bibinfo{author}{R{\'e}nyi, A.}
\newblock \bibinfo{title}{On the evolution of random graphs}.
\newblock \emph{\bibinfo{journal}{Publ. Math. Inst. Hung. Acad. Sci}}
  \textbf{\bibinfo{volume}{5}}, \bibinfo{pages}{17--60} (\bibinfo{year}{1960}).

\bibitem{kermack1927contribution}
\bibinfo{author}{Kermack, W.~O.} \& \bibinfo{author}{McKendrick, A.~G.}
\newblock \bibinfo{title}{A contribution to the mathematical theory of
  epidemics}.
\newblock In \emph{\bibinfo{booktitle}{Proceedings of the Royal Society of
  London A: mathematical, physical and engineering sciences}}, vol.
  \bibinfo{volume}{115}, \bibinfo{pages}{700--721} (\bibinfo{organization}{The
  Royal Society}, \bibinfo{year}{1927}).

\bibitem{anderson1992infectious}
\bibinfo{author}{Anderson, R.~M.}, \bibinfo{author}{May, R.~M.} \&
  \bibinfo{author}{Anderson, B.}
\newblock \emph{\bibinfo{title}{Infectious diseases of humans: dynamics and
  control}}, vol.~\bibinfo{volume}{28} (\bibinfo{publisher}{Wiley Online
  Library}, \bibinfo{year}{1992}).

\bibitem{Starnini2013Immunization}
\bibinfo{author}{Starnini, M.}, \bibinfo{author}{Machens, A.},
  \bibinfo{author}{Cattuto, C.}, \bibinfo{author}{Barrat, A.} \&
  \bibinfo{author}{Pastor-Satorras, R.}
\newblock \bibinfo{title}{Immunization strategies for epidemic processes in
  time-varying contact networks.}
\newblock \emph{\bibinfo{journal}{Journal of Theoretical Biology}}
  \textbf{\bibinfo{volume}{337}}, \bibinfo{pages}{89--100}
  (\bibinfo{year}{2013}).

\bibitem{dunbar2009social}
\bibinfo{author}{Dunbar, R.~I.}
\newblock \bibinfo{title}{The social brain hypothesis and its implications for
  social evolution}.
\newblock \emph{\bibinfo{journal}{Ann. Hum. Biol.}}
  \textbf{\bibinfo{volume}{36}}, \bibinfo{pages}{562--572}
  (\bibinfo{year}{2009}).

\bibitem{tamarit2018cognitive}
\bibinfo{author}{Tamarit, I.}, \bibinfo{author}{Cuesta, J.~A.},
  \bibinfo{author}{Dunbar, R.~I.} \& \bibinfo{author}{S{\'a}nchez, A.}
\newblock \bibinfo{title}{Cognitive resource allocation determines the
  organization of personal networks}.
\newblock \emph{\bibinfo{journal}{Proc. Natl. Acad. Sci. USA}}
  \textbf{\bibinfo{volume}{115}}, \bibinfo{pages}{8316--8321}
  (\bibinfo{year}{2018}).

\bibitem{dunbar2018social}
\bibinfo{author}{Dunbar, R.}
\newblock \bibinfo{title}{Social structure as a strategy to mitigate the costs
  of group living: a comparison of gelada and guereza monkeys}.
\newblock \emph{\bibinfo{journal}{Anim. Behav.}}
  \textbf{\bibinfo{volume}{136}}, \bibinfo{pages}{53--64}
  (\bibinfo{year}{2018}).

\bibitem{Rocha2010Information}
\bibinfo{author}{Rocha, L. E.~C.}, \bibinfo{author}{Liljeros, F.} \&
  \bibinfo{author}{Holme, P.}
\newblock \bibinfo{title}{Information dynamics shape the sexual networks of
  internet-mediated prostitution}.
\newblock \emph{\bibinfo{journal}{Proc. Natl. Acad. Sci. USA}}
  \textbf{\bibinfo{volume}{107}}, \bibinfo{pages}{5706--5711}
  (\bibinfo{year}{2010}).

\bibitem{rocha2011simulated}
\bibinfo{author}{Rocha, L.~E.}, \bibinfo{author}{Liljeros, F.} \&
  \bibinfo{author}{Holme, P.}
\newblock \bibinfo{title}{Simulated epidemics in an empirical spatiotemporal
  network of 50,185 sexual contacts}.
\newblock \emph{\bibinfo{journal}{PLoS Comput. Biol.}}
  \textbf{\bibinfo{volume}{7}}, \bibinfo{pages}{e1001109}
  (\bibinfo{year}{2011}).

\bibitem{Stehl2011High}
\bibinfo{author}{Stehl\'{e}, J.} \emph{et~al.}
\newblock \bibinfo{title}{High-resolution measurements of face-to-face contact
  patterns in a primary school}.
\newblock \emph{\bibinfo{journal}{PLoS ONE}} \textbf{\bibinfo{volume}{6}},
  \bibinfo{pages}{23176} (\bibinfo{year}{2011}).

\bibitem{Holme2012Temporal}
\bibinfo{author}{Holme, P.} \& \bibinfo{author}{Saram\"{a}ki, J.}
\newblock \bibinfo{title}{Temporal networks}.
\newblock \emph{\bibinfo{journal}{Phys. Rep.}} \textbf{\bibinfo{volume}{519}},
  \bibinfo{pages}{97--125} (\bibinfo{year}{2011}).

\bibitem{sulo2010meaningful}
\bibinfo{author}{Sulo, R.}, \bibinfo{author}{Berger-Wolf, T.} \&
  \bibinfo{author}{Grossman, R.}
\newblock \bibinfo{title}{Meaningful selection of temporal resolution for
  dynamic networks}.
\newblock In \emph{\bibinfo{booktitle}{Proceedings of the Eighth Workshop on
  Mining and Learning with Graphs}}, \bibinfo{pages}{127--136}
  (\bibinfo{organization}{ACM}, \bibinfo{year}{2010}).

\bibitem{caceres2011temporal}
\bibinfo{author}{Caceres, R.~S.}, \bibinfo{author}{Berger-Wolf, T.} \&
  \bibinfo{author}{Grossman, R.}
\newblock \bibinfo{title}{Temporal scale of processes in dynamic networks}.
\newblock In \emph{\bibinfo{booktitle}{Data Mining Workshops (ICDMW), 2011 IEEE
  11th International Conference on}}, \bibinfo{pages}{925--932}
  (\bibinfo{organization}{IEEE}, \bibinfo{year}{2011}).

\end{thebibliography}

%


\end{document}